\documentclass[preprint,12pt,3p]{elsarticle}

\usepackage{amssymb}
\usepackage{lineno,hyperref,graphicx,amsmath,color,xcolor}
\usepackage{multirow, tabularx,xfrac}

\graphicspath{{figures/}}
\newcolumntype{Y}{>{\centering\arraybackslash}X}

\journal{Ocean Modelling}

\newcommand{\eng}[1]{{#1}}
\newcommand{\ansnew}[1]{{#1}}
\begin{document}

\begin{frontmatter}

\title{Adaptation of NEMO-LIM3 model for multigrid high resolution Arctic simulation}

\author[label1]{Alexander Hvatov\corref{cor1}}
\ead{matematik@student.su}
\author[label1]{Nikolay O. Nikitin\corref{cor1}}
\ead{nikolay.o.nikitin@gmail.com}
\author[label1]{Anna V. Kalyuzhnaya}
\ead{kalyuzhnaya.ann@gmail.com}
\author[label2]{Sergey S. Kosukhin}
\ead{skosukhin@gmail.com}
\address[label1]{ITMO University, 197101, 49 Kronverksky pr., St Petersburg, Russia}
\address[label2]{Max Planck Institute for Meteorology, 20146, Bundesstr. 53, Hamburg, Germany}

\cortext[cor1]{Corresponding author}

\begin{abstract}

\ansnew{High-resolution regional hindcasting of ocean and sea ice plays an important role in the assessment of shipping and operational risks in the Arctic Ocean. The ice-ocean model NEMO-LIM3 was modified to improve its simulation quality for appropriate spatio-temporal resolutions. A multigrid model setup with connected coarse- (14 km) and fine-resolution (5 km) model configurations was devised. These two configurations were implemented and run separately. The resulting computational cost was lower when compared to that of the built-in AGRIF nesting system. Ice and tracer boundary-condition schemes were modified to achieve the correct interaction between coarse- and fine grids through a long ice-covered open boundary. An ice-restoring scheme was implemented to reduce spin-up time. The NEMO-LIM3 configuration described in this article provides more flexible and customisable tools for high-resolution regional Arctic simulations.}

\end{abstract}


\end{frontmatter}


\section{Introduction}
\label{sec1}

\ansnew{High-resolution regional models configured for specific target domains are widely used to reproduce metocean extremes reliably \cite{chan2014value}. Due to the lack of observation data, they also play a critical role in the assessment of risks for offshore development and shipping \cite{ehlers2014scenario} in the Arctic Ocean. The existing statistical risk-assessment methods require a hindcast of the essential ocean and ice variables --- e.g., sea surface height, currents velocity, ice concentration/thickness distributions and drift velocity --- with a considerable time coverage of 30 years or more \cite{lopatoukhin2009multivariable}.

Modern ocean-ice reanalyses provide useful and quite reliable information on the physical state of the Arctic Ocean in the last 70 years \cite{Uotila2018,chevallier2017intercomparison}. Unfortunately, most reanalyses are still conducted at low time resolutions, usually providing only daily-mean fields, so that the hourly variability of ocean dynamics remains uncaptured. Moreover, the historical coverage of different reanalyses is not homogeneous. For these reasons, determining the most suitable reanalyses for a statistical estimation of extreme events is a difficult task. Ideally, such reanalysis data should cover long time periods, have high spatial resolution and contain data for different depths.

The quality of extreme-event simulations has been found to be sensitive to the horizontal resolution of a model \cite{jianping2007effects}, especially in coastal areas with abrupt bathymetry changes. Given the conditions outlined above, we aimed to create a dataset with appropriate historical coverage and sufficiently high spatial resolution.

The computational cost of performing global high-resolution ocean simulations (i.e.  less or equal to 5 km) remains prohibitive, at least if the simulation is to be concluded within reasonable time constraints. Given the spatial domain of interest in this article consists of the Russian Arctic seas, we decided to base our research on the regional high-resolution ice-ocean simulations and the expertise of the associated community, as opposed to well-known global configurations like ORCA025 \cite{bourdalle2006climatology}.

There are various projects in the field of regional-Arctic modelling. Multi-purpose high-resolution models are often based on regional or nested configurations. For example, the nested Arctic-FVCOM model \cite{ArcticFVCOM} reproduces shallow-water transport, the regional NEMO-based configuration RASM \cite{RASM} provides detailed atmospheric circulation, and the CREG project \cite{dupont_ice_model} produces detailed short-term ocean predictions for hazard warning.

We reached our goals with a custom simulation framework comprising three main components: the WRF atmospheric model \cite{powers2017weather}, the WaveWatch III spectral wave model \cite{tolman2009user} and the NEMO ocean model, the last one coupled with the LIM3 ice model. Together, these models allowed us to obtain the desired long-term datasets.

Domain-specific observational-driven post-correction described in \cite{gusarov2017spatially} was applied for the atmospheric model output, but no modifications were made in the source code of either WRF or WaveWatch.

In the ocean-related part of the simulation framework, the two regional configurations run disjointly on the corresponding different scales to improve computational performance. Output data from a completed low-resolution run are used as input data for a high-resolution run.

Regional Arctic modelling poses several challenges for which adequate solutions are still lacking. In particular, long open boundaries remain an ill-posed problem \cite{OBCillposed} and a source of instability, especially for our open ice-boundary case. One way to circumvent this difficulty is to place the boundary as far from the region of interest as possible. This approach can only be used when there is no constraint in computational cost, and therefore was not an alternative we could contemplate.

In this article, we discuss particular solutions to problems associated with building a high-resolution regional NEMO-LIM3 configuration, the issues we faced in adapting NEMO configuration and the validation of the implemented 
modifications. Special attention is given to the long ice-covered open-boundary problem and indirect ice-data assimilation. The proposed solutions are not restricted to the specific Arctic region
used in our case-study.}
The modified source code is publicly available \cite{nemo-multigrid}.

The article is organised as follows. Sec.\ref{sec2} contains detailed descriptions of a particular Arctic modelling problem and of the corresponding computational model. Sec.\ref{sec3} describes the grid generation problem and the proposed solutions. Sec.\ref{sec4} covers long open-boundary condition problems, in particular those with high ocean-tracer and ice-exchange ratios. Sec.\ref{sec5} is dedicated to the ocean-ice spin-up and restoring
procedures.

\section{Model description}
\label{sec2}

\ansnew{
We now describe the experimental configuration we built with the specific purpose of validating the modifications made to the NEMO ocean model setup. This configuration is similar to an actual long-term hindcasting system: it starts with a ``climatological" (i.e. cyclic) spin-up with data corresponding to the year 2013. Once a fully operational system is obtained, it can be used in longer-term simulations: our future target is a reliable hindcast over a 50 years ago from now.

According to \cite{Uotila2018}, 
mapping data from low- to high-resolution grids can lead to reduced quality,
in particular at external boundaries. Moreover, structural problems of the
input data --- e.g.  insufficient historical coverage and time resolution of 
different reanalyses, low spatial resolution of ice climatological 
databases --- can also be exacerbated through interpolation into higher
resolutions.
To minimize these effects, the simulation workflow comprises two stages, 
one with higher resolution than the other. We refer to these throughout
this article as coarse- and fine-resolution model configurations. 
Both configurations are based on 
NEMO-LIM3 (version 3.6 STABLE, build 7873) \cite{NemoGuide}.}

\begin{figure} [ht!]
      \center
      \includegraphics [scale=1.3] {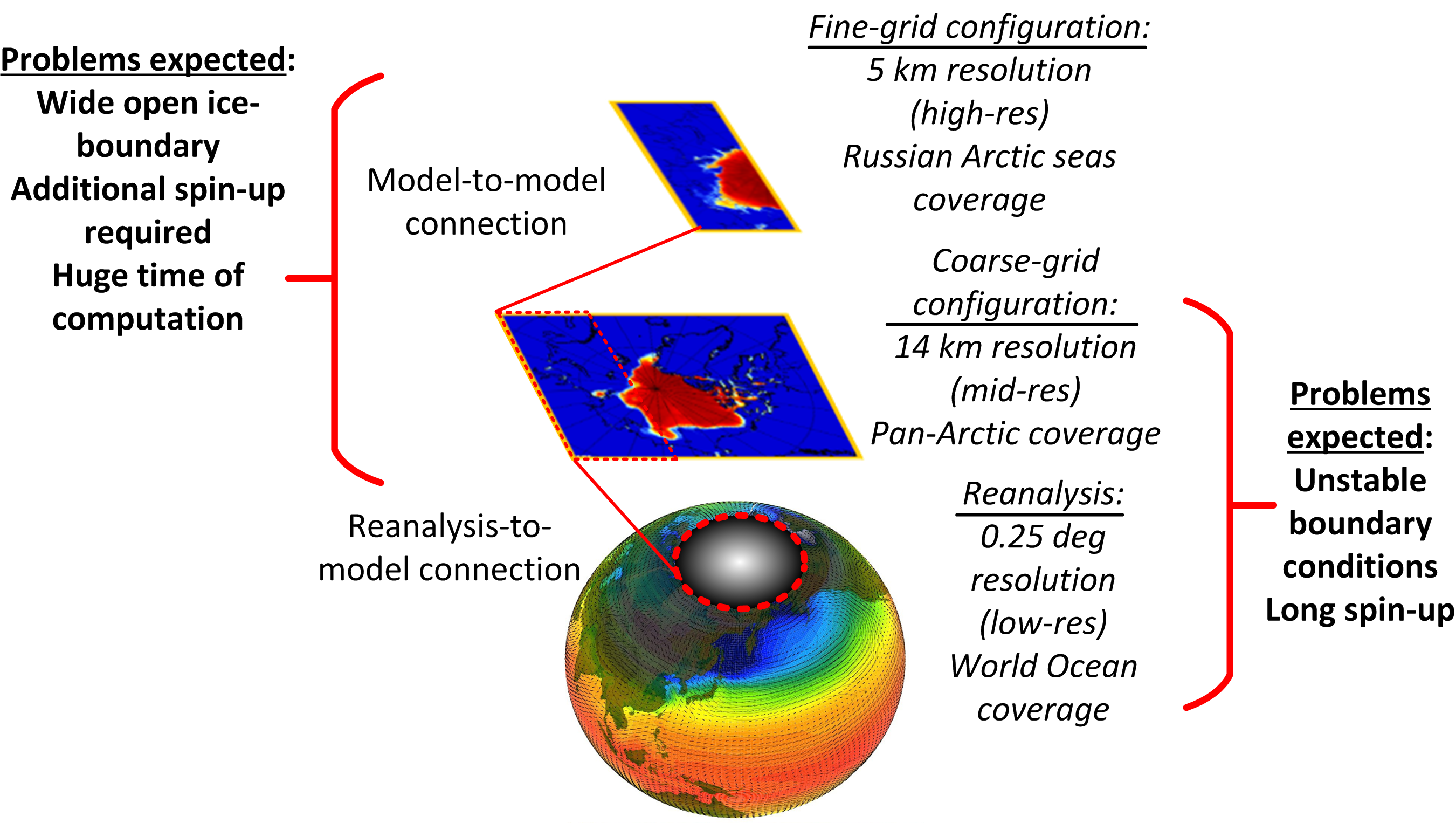}
      \caption{Expected issues in the two-scale simulation workflow. 
      Boundary condition data for the coarse-grid configuration is obtained from global reanalyses. Corresponding data for the downscaled fine-grid configuration is obtained from the output of the preceding stage.}
      \label{fig_issues}
    \end{figure}

Fig.~\ref{fig_issues} depicts the framework just described and outlines
the problems we anticipate as we move 
to the higher-resolution configuration of the Arctic region. 
The first regional model contains the entire Arctic region, whereas the 
second consists of the Russian Arctic seas only.

\begin{figure} [ht!]
  \center
  \includegraphics [width=0.6\linewidth] {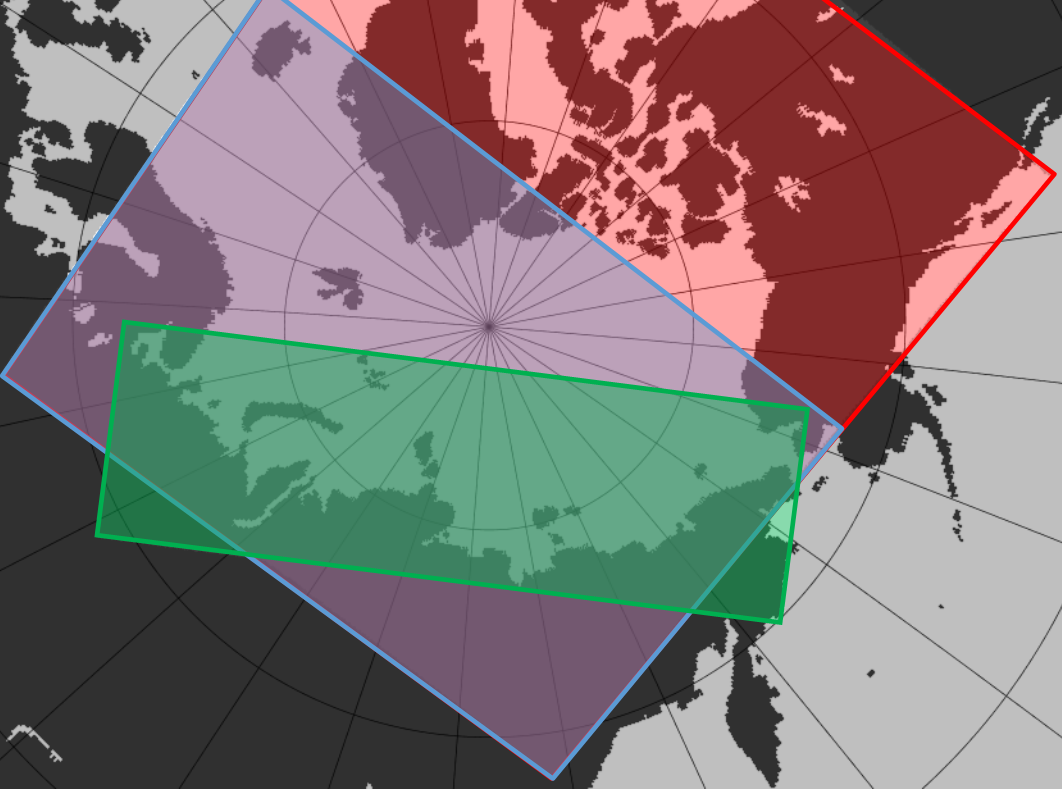}
  \caption{Possible positions of the fine grid with respect to the 
  coarse grid. The coarse grid corresponds to the area covered
  by both red and blue rectangles. The blue rectangle denotes a 
  possible AGRIF-based nesting. The green rectangle corresponds to
  our custom fine grid.}
  \label{fig_grid_comparison}
\end{figure}

The solution illustrated in Fig.~\ref{fig_grid_comparison} consists of two model configurations, each using a custom grid with a different alignment. The coarse-grid configuration is used in a spin-up procedure 
for generating initial and boundary conditions for the fine-grid configuration.
It provides `rough' values for regional ocean- and ice-parameter fields. These configurations are coupled as follows: 
data from the low-resolution grid are transferred to the fine grid at long simulation-time intervals, e.g. yearly.
These are downscaled (interpolated) once the corresponding low-resolution dataset is available. The fine-grid configuration computes finely resolved fields for the desired target region --- the Russian Artic seas, in this case.
This approach allows for a flexible management of computation costs and increased
stability of the fine-grid configuration, with the minor drawback of a
relatively short additional spin-up time.

\ansnew{The actual input data used in our runs were collected from various sources. The atmospheric forcing for the NEMO configurations, both of which used the CORE bulk module parametrisation \cite{NemoGuide}, was obtained from the Polar WRF model \cite{powers2017weather} in the simulation framework. 

WRF forcing was computed for the same time period covered by the NEMO runs. In order to better adapt the WRF model to Arctic conditions, the Pleim-Xiu Land Surface Model and Asymmetric Convective Model 
parameterisations were activated \cite{krieger2009p1}. Because WRF can only be used with its self-generated grid --- 
in this case, a pan-Arctic 14 km grid --- we generated a 480$\times$440 grid as close as possible to that of the coarse grid (see Sec.\ref{sec3}). 

Input variables and initial conditions for WRF model --- e.g. ice cover, surface temperature, 2D and 3D atmospheric parameters --- were obtained from the OSI SAF dataset \cite{OSISAFreference} (ice cover from 1979), the ERA-Interim \cite{dee2011era} (surface temperature, atmospheric parameters from 1979) and NCEP-DOE 2 reanalysis \cite{kanamitsu2002ncep} (all variables for years before 1979). 

The output WRF fields were then interpolated to both coarse- and fine grids to provide applicable data. Although NEMO provides a built-in interpolation feature, we chose to use the external tool YAGO \cite{yago} in the interest
of transparency. The WRF output variables used as surface forcing in NEMO --- namely U and V wind components at 10m, air temperature at 2m and mean sea level pressure --- were stored at an 1h time resolution.

The NEMO with CORE bulk parametrisation requires other input variables still:
short- and long-wave radiations, humidity, precipitation, snow.
These were obtained from a
DFS 5.2 reanalysis \cite{dussin2016making} in 24h time resolution. 

Nevertheless, the source of NEMO's atmospheric-forcing data are not 
a crucial point for the experiments performed in Sec.\ref{sec4}-\ref{sec5}.
The validation presented in these sections is based on the 
the comparison of the results obtained with the original and modified NEMO 
configurations, both driven by the same atmospheric forcing. Any bias or
inaccuracy in these input data would be manifest in both outputs.

Other input variables and initial conditions --- e.g.
ice cover, surface temperature, 2D and 3D atmospheric parameters --- were
obtained from the OSI SAF dataset \cite{OSISAFreference} (ice cover from 1979), 
the ERA-Interim \cite{dee2011era} (surface temperature, atmospheric parameters 
from 1979) and NCEP-DOE 2 reanalysis \cite{kanamitsu2002ncep} (all variables 
for years before 1979). 

Daily ice-concentration data from the OSI SAF dataset \cite{OSISAFreference} and weekly ice-thickness data combined from CryoSat-2 and SMOS satellites \cite{ricker2017weekly} 
were used in the implemented ice-restoring scheme for NEMO model described in Sec.\ref{sec5}.

Data from the GLORYS2V4 global ocean reanalysis \cite{GLOBALreference} was 
used as boundary and initial conditions for the coarse-grid NEMO configuration.
This reanalysis was chosen as a compromise between availability, time coverage, 
and resolution \cite{Uotila2018}.

Initial and boundary conditions for the fine-grid runs were obtained by downscaling (with interpolation) both input and output of coarse-grid simulations and saved as restart files for initial conditions and hourly
input files for boundary conditions for fine-grid simulations
corresponding to the same simulation timeframe.}

\ansnew{
At first glance, this procedure may seem to be redundant, as NEMO provides
built-in nesting through the AGRIF package \cite{AGRIFreference}, which
allows embedding a high-resolution area into a lower-resolution grid
and automatically maps the data across grids. 
Unfortunately, the mechanism implemented in the AGRIF package is not 
satisfactory for our purposes: using this package would require us
to perform our simulations over areas
outside of the domain of interest (as shown in Fig.~\ref{fig_grid_comparison}).
In addition, AGRIF requires low- and high-resolution computations to
be carried out concurrently in connected resources, while we aim to have 
freedom to schedule the low- and high-resolution runs on different 
resources, fully independently.
Finally, the actual implementation of the AGRIF package available 
in the stable build of NEMO 3.6 cannot be used with the coupled LIM3-model,
since it supports LIM2 only.

To select appropriate grid resolutions for both coarse- and fine-grid
configurations, 
we conducted a set of preliminary experiments using 250 computational 
nodes of the Lomonosov-II supercomputer \cite{L2ref}. 
In these experiments, simulations of the fine-grid domain
(the Russian Arctic seas) with a 2-kilometer grid (2750$\times$1000) 
took 173 hours wall-time per simulated year ($\approx$ 1800 node days), 
whereas a 3.5-kilometer 
grid (1571x571) took 50 hours wall-time per simulated 
year ($\approx$ 520 node days).
These estimations are too high for either the experimental or real-world
applications.
For the 5-kilometer grid (1100$\times$400) the simulation took 18 hours 
wall-time per simulated year ($\approx 185$ node days), 
which makes this resolution a viable choice. }

\ansnew{
Even if the previously mentioned restrictions could be ignored, the AGRIF-based solution would lead to an increase in computation time of up to 50\% when downscaling. This is a consequence of nested grid positions being restricted by the parent grid alignment in AGRIF, as shown in Fig.~\ref{fig_grid_comparison}.

Therefore, the coarse (14 km) grid configuration was separated from fine (5 km) configuration. The resolution is selected as a compromise between the fine-grid and the reanalysis (0.25 degree) resolutions.

The baroclinic timestep (rdt) was taken as $960$ and $240$ for the 14 km and 5 km resolutions, respectively. Both configurations have 18 vertical levels. The vertical grid is configured \cite{NemoGuide} such that a sparse step
near the bottom is achieved. These specific values are not critical for the modifications to the NEMO model proposed herein, and are provided only to fully describe the experimental configuration used in the validation process.
}

\begin{figure} [ht!]
      \center
      \includegraphics [scale=1.0] {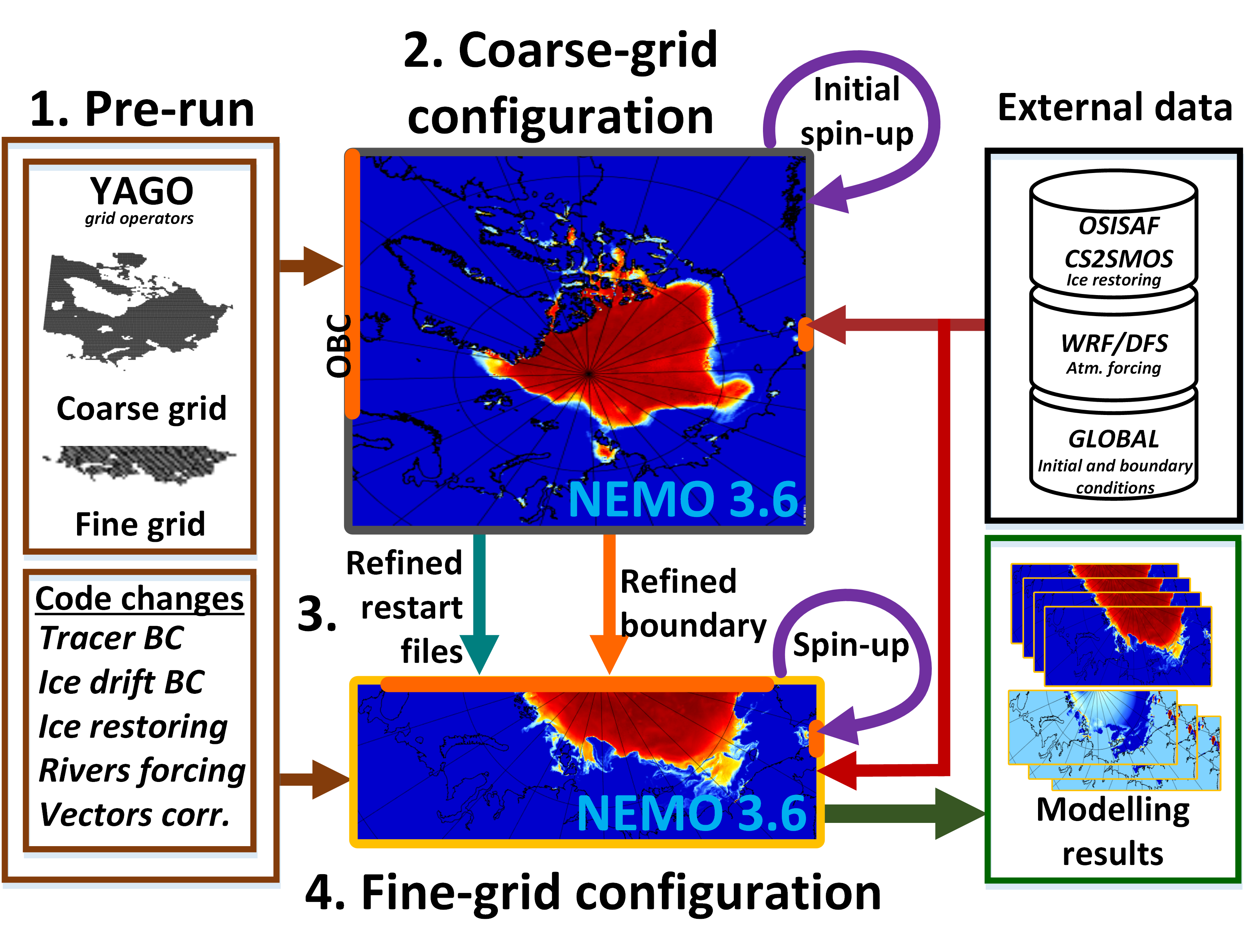}
      \caption{Common scheme for the two-grid coupled modelling system. 
      The four stages are presented: pre-processing, two stages of 
      regional simulation and the intermediate inter-grid refinement stage.}
      \label{figsystem}
    \end{figure}

Fig.~\ref{figsystem} illustrates 
the construction and operation of the simulation framework
described in this article and outlines four main stages:
\ansnew{(i) grid preparation, model source-code 
modification (\ref{App_code_modifications} provides a full list
of modifications and associated source files),
pre-processing,} 
(ii) computation with the low-resolution configuration, 
(iii) extraction and refinement of boundary and initial conditions for the 
high-resolution computation and 
(iv) computation with the high-resolution configuration. 

\section{Grid generation and adjustment}
\label{sec3}

NEMO employs an Arakawa Staggered C-Grid generalised 
to three dimensions. 
Apart from a few simple analytical horizontal grids,
such as regular latitude-longitude grid, {\it f}\.-plane and $\beta$-plane, 
the model also supports general curvilinear orthogonal grids.

Latitude-longitude grids are unsuitable for polar regions because of the 
extreme cell-size variation. A (near) constant cell size increases model stability and decreases computational costs, as the admissible model time step is related to the 
smallest cell size. To avoid this caveat, we generate a curvilinear grid based on a projection with adjusted parameters. This grid  must be provided as four sets of geographical 
coordinates and scale factors, each associated with one type of grid point:
the centres of two-dimensional cells (t-points), the centres of their edges (u- and v-points), and their vertices (f-points) \cite{NemoGuide}. 

We developed a custom optimisation tool to compute grid parameters, YAGO \cite{yago}, which is publicly available. The parameters to be optimised are the geographical coordinates of the projection's centre and the associated affine transformations between the coordinate spaces, i.e. the optimisation target function is the minimisation of associated projection distortions. Using this tool and methodology, we analysed several types of projections. The corresponding distortion values are shown in \ref{App_grids}. Due to the low aspect ratio of the computational domain, the best result was obtained with the Stereographic projection, which we therefore used to generate the model grid.

The obtained grid data were converted to the format required by NEMO \cite{NemoGuide}. However, these grids are not model-specific, and can be used in other ocean models, as well as other interconnected dynamic models (e.g., for atmosphere, waves) for the same domain.

\ansnew{Using the curvilinear grid introduces one additional difficulty. Most reanalysis data are available in the Eastward-Northward coordinate  system, also known as geographical notation, in which components of vector variables (currents, wind, drift) are given in terms of unit vectors in the East and North directions. The same convention is usually assumed by
visualization tools. NEMO, on the other hand, defines vector-variable components in terms of the unit vectors aligned with the local grid axes (local notation). This problem is partially solved by NEMO's SBC module, which implements pairwise rotation from E-N space to U-V space (the \textit{rotation pairing} column in \textit{namelist\_cfg}). However, this functionality is available only for input vector variables (e.g., wind data) : the unstructured boundary-condition files for barotropic and baroclinic currents cannot be rotated this way in the standard version of NEMO. A similar situation arises for NEMO output data, which are also provided in local grid coordinates and should be post-processed manually. Finally, since the metadata of NetCDF usually does not clearly specify in which coordinate system the data are given, all these issues are made further error-prone.

To better handle the mapping between the two coordinate systems, we introduced changes to NEMO's source code. These implement the necessary transformations between local and geographical notations at both input and output stages. This allows all datasets to be pre- and post-processed consistently in the unified geographical notation, keeping the internal
representation in local notation independent and automated. Details on the modifications are presented in \ref{App_vectors}.}

\section{Open boundary conditions}
\label{sec4}

\ansnew{Open boundary conditions (OBC) are an essential part of the configuration of regional hydrodynamic models, since they play a major role in defining the structures of thermohaline circulation and currents. We faced several OBC-related issues while setting up the regional configuration of NEMO. For instance, tracer gradients may vary rapidly at long open boundaries, leading to numerical artefacts in heat (and salt) fluxes at the boundary. The result is that inflow and outflow may become inconsistent, and this inconsistency can cause stability problems, which become increasingly manifest the longer the boundary is.}

Another problem arises at ice-covered open boundaries, which require ice
concentration, drift, ice- and snow thickness as input data. 
However, external data for the ice boundary are only available in low 
resolution, if at all.
The commonly adopted approach to circumvent this problem is similar
to the closed-wall boundary condition used in \cite{dupont_ice_model}.

As an alternative, we implemented a data-driven OBC scheme through an ice-covered 
boundary in NEMO-LIM3 \cite{lim36refguide}.
The ice-drift conditions used in the unmodified NEMO 
scheme refer to ocean current velocities instead of the actual drift velocity data, 
which negatively affects the quality of the fine-grid simulation.

This section is organised as follows: 
summaries of possible long open-boundary treatments
for tracer OBC and for sea-ice drift OBC are presented
in Secs.\ref{sec41} and \ref{sec42}, respectively.
The corresponding mathematical formulation is further detailed in
\ref{App_tracer_boundary}.
Experimental results used in the validation of the modifications made to the
boundary conditions in NEMO are presented in Sec.\ref{sec43}.

\subsection{Tracer boundary conditions}
\label{sec41}

Open boundary conditions for regional configurations of ocean 
models are a topic of many research, and there seems to be no simple 
answer as 
to which scheme performs better for any given case \cite{OBCdiscussion}. 

At the tracer open boundary, the sign of a tracer's phase velocity  determines 
whether it is an inflow or an outflow, each case being treated differently. 
For inflow regions, radiation conditions are applied: these take into account 
the physical properties of heat radiation. For outflow regions, 
nudging is applied instead. 
The corresponding schemes available in NEMO are the Orlanski radiation scheme
(for inflow) and the Flow Relaxation Scheme (FRS) (for nudging/outflow).
Further detail is given in \ref{App_tracer_boundary}.

\begin{figure} [ht!]
  \center
  \includegraphics [scale=0.4] {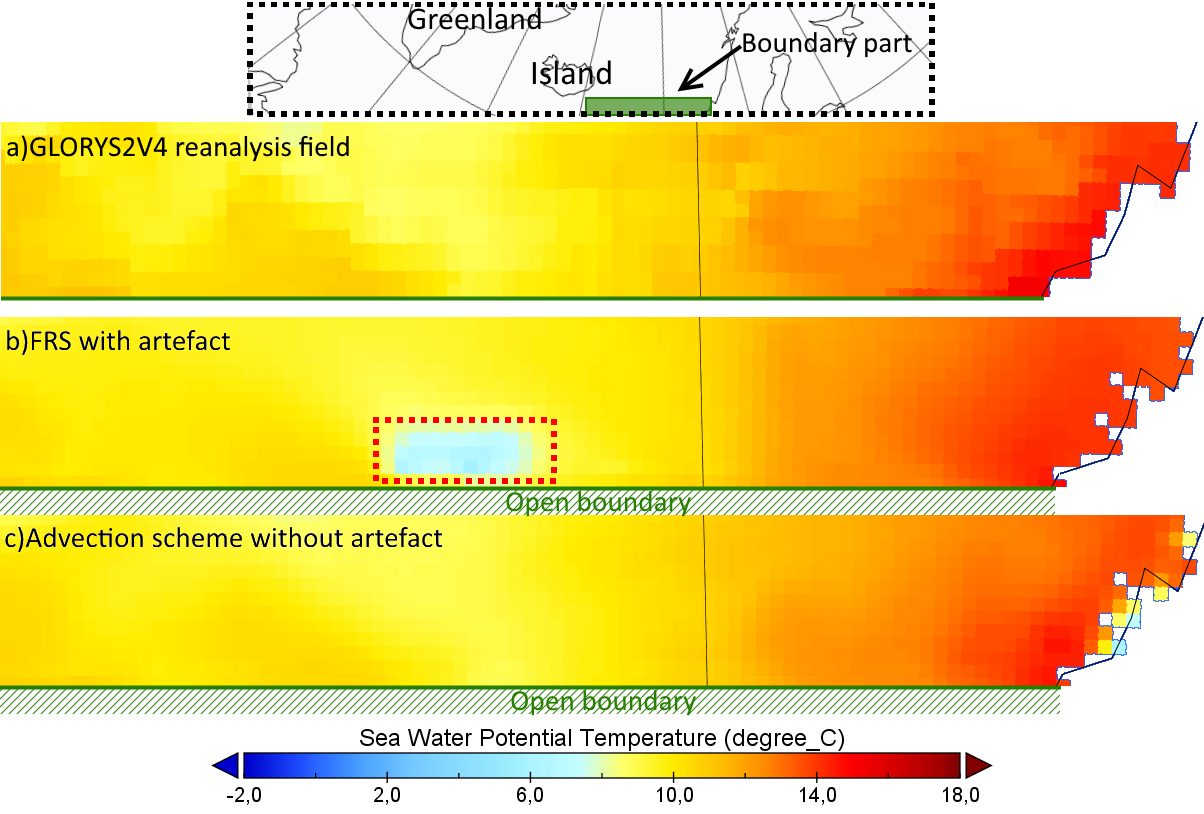}
  \caption{\eng{Affected temperature field} for 15.09.2013: 
  (a) GLORYS2V4 global reanalysis (no artefacts), 
    (b) NEMO with FRS boundary conditions (open-boundary artefacts, dashed 
  frame), 
  (c) NEMO with ADV+FRS boundary conditions (no open-boundary artefacts). 
  \ansnew{The green zone at the map is an analyzed section of the boundary.}}
  \label{fig_artefacts}
\end{figure}

In the long open-boundary case, there can be multiple inflow and outflow zones along the boundary, and the correspondingly different tracer behaviours can become significant. Quantitatively large changes imposed on tracer values at the boundaries by either radiation boundary conditions (for the inflow) or nudging (for the outflow) can lead to non-physical values, as neighbouring cells display alternating flux signs. The detrimental effect of these artefacts is greater when the zone of interest is close to the boundary. In addition, post-processing must be used
to filter these artefacts from the output data. Fig.~\ref{fig_artefacts}b shows an example of such artefact in a 
a section of the wide boundary of the coarse grid.

In order to prevent these artefacts , we replaced the 
inflow part of the Orlanski scheme provided in NEMO with the FRS scheme and 
added an ocean flow velocity term in the tracer phase speed.
Using the FRS scheme for inflow and ADV for 
outflow follows from work presented in \cite{OBCreference}.

\subsection{Ice-drift boundary conditions}
\label{sec42}

\ansnew{Open boundary conditions for sea-ice dynamics are not well 
described in the literature. 
Some implementations of ice boundaries can be found
in \cite{lim36refguide}, but the references therein only consider scalar 
variables like ice concentration and ice/snow thickness. 
Existing references generally provide no treatment for sea-ice drift, 
recommending that long ice-covered boundaries be avoided.}
Thus, implementing a regional model with such boundary
conditions is still a research challenge.

The limitations of open boundary implementations for sea ice are 
closely related to the quality issues of external sea-ice datasets. 
Usually, ice concentration, thickness and drift fields are not available in high spatial resolution.
Additionally, ice thickness and ice drift are available  with restricted historical coverage. In case of ice thickness there exist seasonal restrictions as well.

In the satellite observations datasets we used, ice drift is computed 
in Lagrangian coordinates \cite{SatelliteDrift}
as the displacement of characteristic points (e.g. ice edges).
Then it has to be mapped to Eulerian
coordinates before it can be used as input for NEMO \cite{NemoGuide} or compared
to its output.

\ansnew{The two-scale (i.e. coarse- and fine configurations) approach described in 
this article avoids most problems related to ice boundary data. 
The fine grid has a long ice-covered boundary, but the required input
data are obtained from coarse simulation in an appropriate spatio-temporal 
resolution. Moreover, using the same tool for both resolutions means that
no additional treatment is required when bridging the data between the
two scales \cite{lim36refguide}.}

The default LIM3 implementation applies
the following rule to ice-drift boundary conditions:
ice-drift is equal to the adjacent grid cell value if it is not ice-free, 
otherwise it is equal to the ocean-current velocity. 
In this way, ice drift is determined by ocean fluxes and wind. 
This is somewhat similar to the free ice-drift conditions proposed 
in \cite{freeicedrift} for the VP (visco-plastic) ice model.
Unfortunately, this approach does not provide satisfactory results for
near-boundary ice-drift values at
long open boundaries that cross ice-covered areas.

In addition to the preceding rule, we included ice-drift boundary data as
FRS conditions.
This allowed us to achieve more realistic ice-drift velocity fields near the
ice-covered boundary. 
We applied this modification only to the fine-grid configuration, 
since the long 
ice-boundary is a major issue in the fine-grid domain only.
The FRS-based drift conditions are provided as input data for the 
fine-grid runs, and are generated by interpolating the
coarse-grid output data.

\begin{figure} [h!]
  \center
  \includegraphics [width=0.8\linewidth] {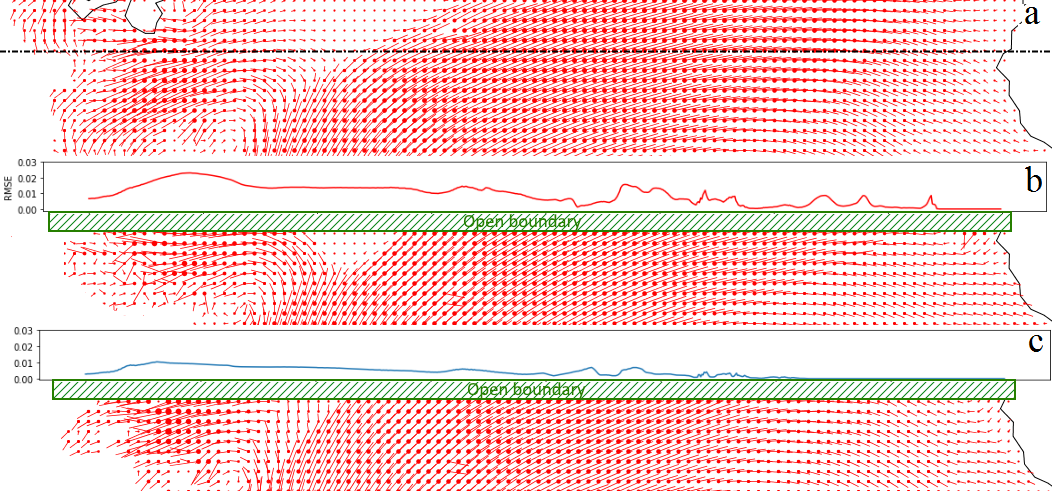}
  \caption{Ice velocity fields near the northern open boundary of the
  fine grid configuration (the dashed line marks the fine configuration grid boundary) (9 Jan 2013): 
  (a) same sub-domain of coarse grid, (b) results of refined grid with default OBC for sea-ice drift 
  (no external drift-values used), (c) implemented OBC for sea-ice drift (including external drift-values). 
  \ansnew{The line plots located above presented fields show the near-boundary 
  ice drift velocity RMSE of the refined model against the coarse grid 
  results.}}
  \label{fig_drift_art}
\end{figure}

Using the coarse-grid ice velocity field as reference
(Fig.~\ref{fig_drift_art}a), 
we assess the quality of the corresponding fine-grid results.
Fig.~\ref{fig_drift_art}b and Fig.~\ref{fig_drift_art}c
correspond to the fields generated using
unmodified and modified sea-ice drift OBC, respectively.
The results obtained with the method proposed herein
display a lower root mean-square error (RMSE)
than those obtained with the unmodified OBC.

This problem is manifest only in the fine-grid configuration, since
the (long) Atlantic boundary is far removed from ice-covered regions
and the Bering strait is quite short
(see Fig.~\ref{fig_grid_comparison}).

\subsection{Validation of the modified open boundary conditions}
\label{sec43}

\subsubsection*{Tracer open boundary conditions}

\begin{figure} [ht!]
  \center
  \includegraphics [width=1\linewidth] {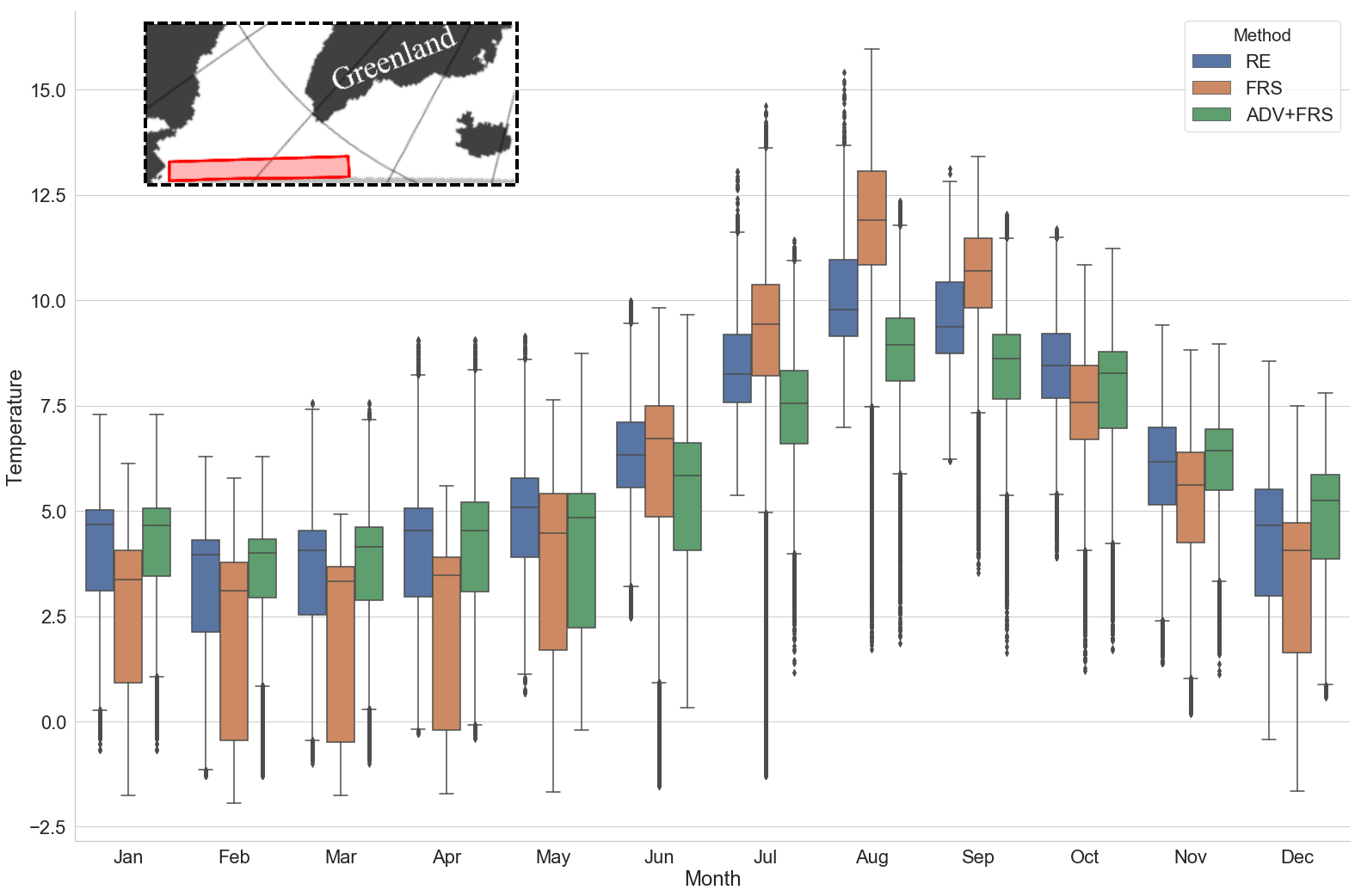}
  \caption{Monthly average temperature boxplot 
  (whiskers - 5$\%$ and 95$\%$ quantiles) for part of the coarse-grid
  domain (top-left corner).
  Blue - GLORYS2V4 global reanalysis (RE) data, 
  orange - default FRS conditions, 
  green - implemented ADV scheme.}
  \label{fig3}
\end{figure}

Fig.~\ref{fig3} shows whisker plots for the temperature values of
the reanalysis data alongside those generated by our runs
using different treatments of boundary conditions.
The FRS scheme produces a larger number of outliers, which we in turn
found to be correlated to temperature-field artefacts. 
Because the monthly-mean data averaged over the relaxation zone
are not smooth, the output must be further corrected in post-processing.

\begin{figure} [ht!]
  \center
  \includegraphics [width=0.9\linewidth] {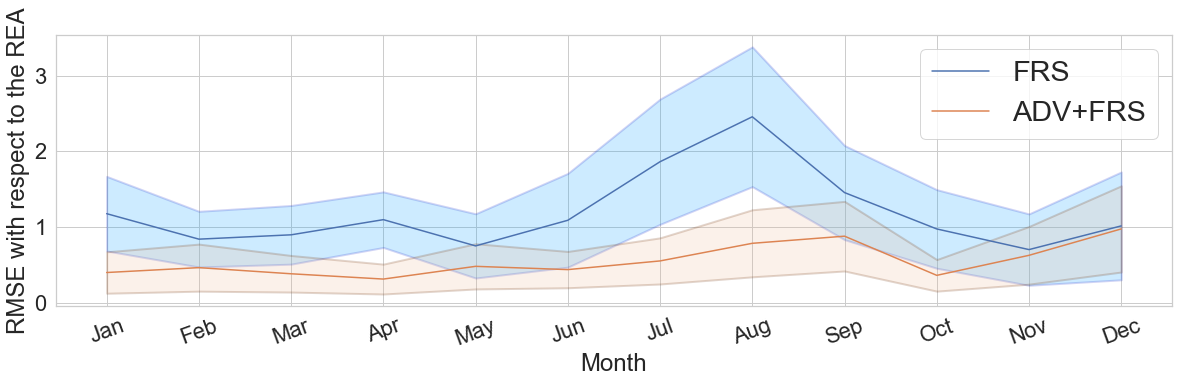}
  \caption{Monthly-averaged temperature RMSE with respect to the GLORYS2V4 
  reanalysis \ansnew{with 95\%-confidence interval}. 
  Blue - NEMO with unmodified FRS OBC, 
  orange - NEMO with modified FRS+ADV OBC.}
  \label{temperature_significance}
\end{figure}

The surface temperature fields obtained using different tracer OBC schemes 
were also compared to the GLORYS2V4 reanalysis. 
We used the RMSE metric with respect to the monthly-average 
reanalysis temperature fields.
The results ---
for the same subdomain of Fig.~\ref{fig3} ---
are shown in Fig.~\ref{temperature_significance}.

For the coarse-grid configuration of NEMO, 
boundary data are obtained exclusively 
from reanalyses. These are often in low resolution only, which can
lead to problems as the data are interpolated.
The ADV tracer scheme, in comparison, performs well even with input
data of lower quality. Because of this increased reliability, 
we opted for adopting this scheme in the coarse-grid configuration.
\ansnew{It is worth mentioning that the simple FRS tracer open-boundary condition could have better efficiency if only the boundary data and the initial fields were already balanced. }

\subsubsection*{Ice drift open boundary conditions}

\begin{figure} [ht!]
  \center
  \includegraphics [width=0.9\linewidth] {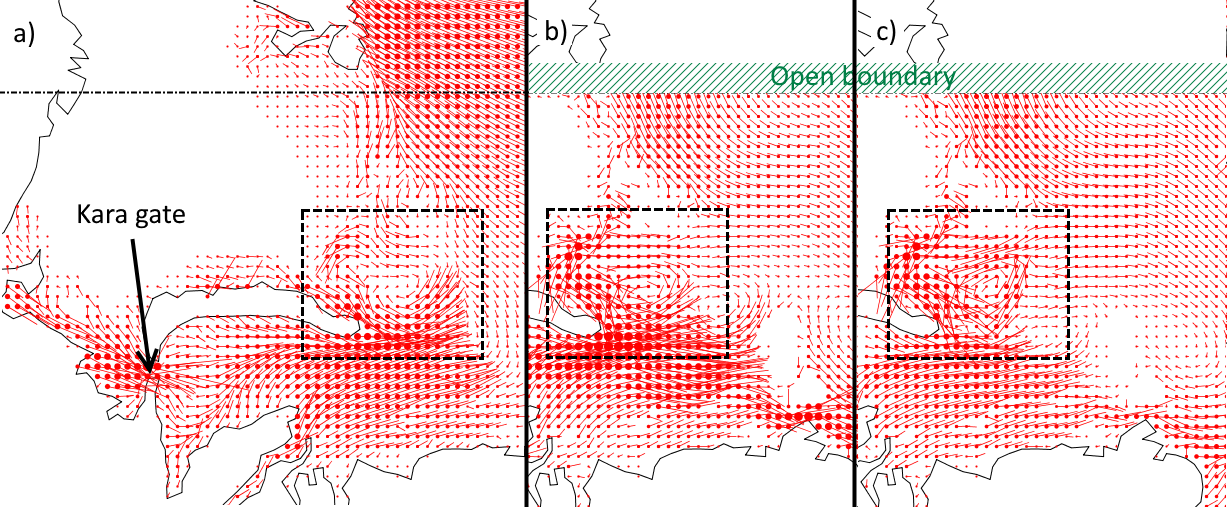}
  \caption{Monthly averaged ice-drift velocity fields (11.2013) on a fine grid configuration subdomain
  (dot-dashed line - fine grid boundary): 
  (a) coarse grid, 
  fine grid:
    (b) default ice drift OBC ,
  (c) implemented ice drift OBC. 
}
  \label{fig_drift}
\end{figure}

Visually comparing (see Fig.~\ref{fig_drift}) the monthly-averaged ice-drift velocity fields generated
with the fine-grid configuration using either the unmodified or modified
OBC scheme shows a
relatively low difference along the boundary between the two cases.

At the same time, the results inside the domain differ substantially,
meaning that different treatments of boundary conditions propagate
into the domain. The non-physical results
outlined in the dashed squares of Fig.~\ref{fig_drift}b can be interpreted as influence of the
ice floes artefacts at the boundary, which  leads to an inadequate ice velocity field.

\begin{figure} [ht!]
  \center
  \includegraphics [width=0.95\linewidth] {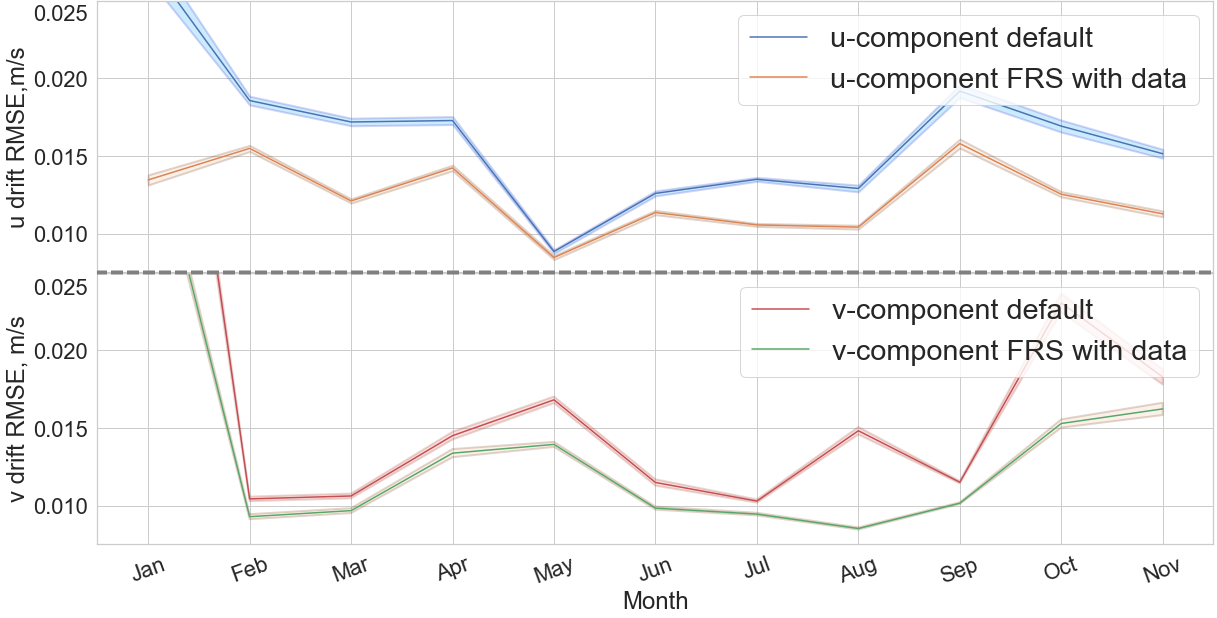}
  \caption{RMSE for one-year runs with different ice drift OBC 
  (entire boundary). Top: drift vector u-component, 
  blue - monthly averaged ice drift field error obtained with default NEMO drift procedure, 
  orange - monthly averaged ice drift field error obtained with implemented FRS scheme with drift boundary data; 
  Bottom: drift vector v-component, 
  red - default OBC , 
  green - implemented FRS scheme.}
  \label{fig_drift_ci}
\end{figure}

Statistical metrics also show that the behaviour of ice on the boundary is more consistent if the default ice-drift boundary conditions of NEMO are augmented with relaxation to the ice-drift boundary data. The comparison shown in Fig.~\ref{fig_drift_ci} is between two model outputs, hence the confidence intervals being so narrow. Errors were computed with respect to the coarse-grid simulation output, where the corresponding velocity values are not affected by boundary conditions and depend only on model parameters.

We have shown that the modified tracer behaviour and 
ice-drift boundary scheme can increase the quality of the model output.
Statistical analysis of the results for near-boundary temperature and 
ice-drift velocity fields confirms the significance of results 
at 95\%-confidence. Therefore, this  modified ice drift OBC scheme
increases the quality of
high-resolution regional Arctic simulations.

\section{Model spin-up}
\label{sec5}

\ansnew{The quality of simulation results is tightly connected to the quality of the initial state. Complex models for which the complete initial state is not known from observational data require a spin-up run to generate a physically consistent first approximation. In order to increase the quality and reduce computational costs for our simulation framework, we conceived a spin-up scheme tailored to our two-scale setup, described in Sec.\ref{sec51}, and implemented additional parameterisations for ice-restoring and river mouth temperature adjustment, described in Secs.
\ref{sec52} and \ref{sec53}, respectively. The validity and effectiveness of this joint approach is presented in Sec.\ref{sec54}.}

\subsection{Ocean-ice spin-up}
\label{sec51}

We extended the usual spin-up procedure for coupled ice-ocean modelling, described in e.g. \cite{BalticIceSpinup}, to work efficiently on our two-grid configuration. This is achieved by performing a primary spin-up run with the coarse grid, and using the resulting state as input for a secondary spin-up with the fine-grid. We then investigate the following two issues. First, whether the spin-up with the fine-grid configuration significantly improves the quality of the fine-grid initial state. Second, whether it is possible to reduce the overall computational cost of the spin-up run.

In common practice, a spin-up procedure consists in running a multi-year simulation without changing boundary conditions and surface forcing from year to year\cite{sevault2009regional}. These simulations are run until quality metrics will be stabilised. We chose to assess the mean absolute error (MAE) and root mean-square error (RMSE) related to our output variables of interest as stopping criterion for the spin-up runs, namely ice concentration and thickness.

Along these lines, we performed ten-year spin-up coarse-model runs using GLORYS2V4 reanalysis data for 2013 as initial conditions for sea temperature and salinity. We also enabled the \textit{ln\_iceini} LIM3 feature to obtain ice-state initial conditions from initial temperature, parameterized with freezing threshold at $2.0^\circ$C, initial concentration 0.9, ice thickness 1 m and snow thickness 0.45 m. After this spin-up, we interpolate the restart files for both ice and ocean to generate input for the fine-grid configuration.

\begin{figure} [ht!]
  \center
  \includegraphics [width=0.8\linewidth] {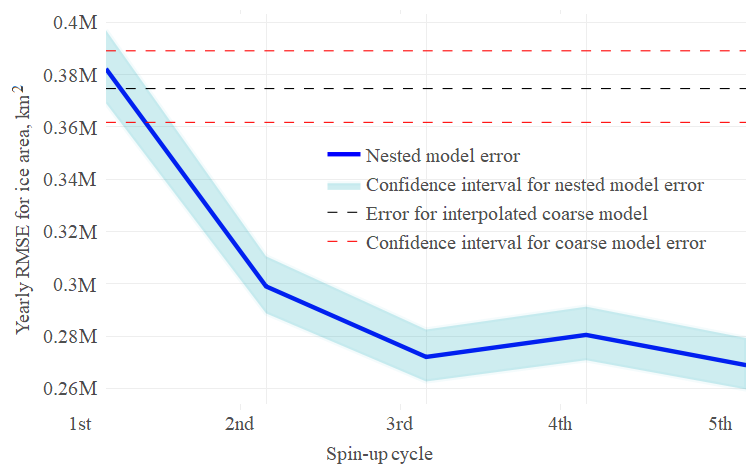}
  \caption{RMSE of ice-covered area for the first five one-year cycles of
  a fine-grid secondary spin-up.
  The black dashed line represents the error associated with the input data
  generated with the coarse-grid spin-up.}
  \label{Fig_SmallScale_SpinUp}
\end{figure}

In addition to the primary coarse-grid configuration spin-up, we then perform a secondary spin-up run on the fine-grid, i.e. we run a series of consecutive one-year fine-grid runs with the aforementioned input and external data for the same year. Fig.~\ref{Fig_SmallScale_SpinUp} shows the errors associated to ice-concentration values for each one-year cycle
of this secondary spin-up. Although the errors for the fine-grid runs are at first relatively high and close to those of the coarse-grid, they decrease after a few cycles. This is to be expected, as the numerical interpolation of fields is not a physical formulation, and the additional fine-grid spin-up improves on the coarse-grid estimate.

These results strongly support using this secondary spin-up procedure on the fine-grid. This approach greatly reduces the computational costs, as the secondary spin-up starts on a much improved initial state than what could be obtained from the reanalysis directly.

\subsection{Ice restoring}
\label{sec52}

The long-term simulation of the coupled ice-ocean system is sensitive to many factors, e.g., model parameterisations or external data for atmospheric forcing and open boundaries. This sensitivity must be taken into account when setting up and calibrating the model, to ensure it produces a stable and physically realistic simulation, hence the careful spin-up procedure previously described. One way to further improve result quality is to assimilate observed values during the simulation\cite{killworth2000effects}. 

\begin{figure} [ht!]
  \center
  \includegraphics [scale=0.3] {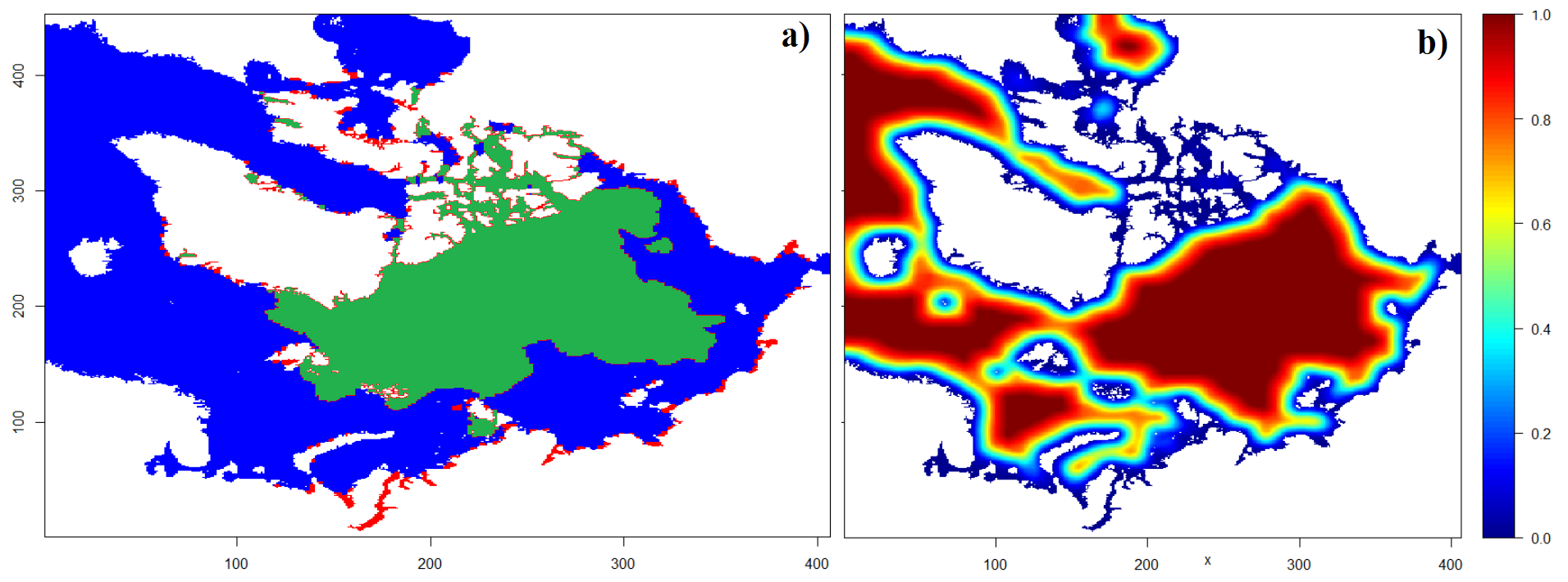}
  \caption{a) Ice cover on the coarse grid from OSI SAF data on 15.09.2013. 
  The green area shows the ice cover confirmed by AARI maps and the red area shows artifacts.
  b) Surface-restoring reliability coefficient~$w$.}
  \label{Fig_anomalies_and_resto}
\end{figure}

Sea-ice concentration data are usually obtained from satellite measurements, as is the case for the OSI SAF dataset which we used. However, such ice-concentration dataset contains artefacts along coastlines due to false concentration values induced by anomalies in brightness temperatures \cite{maslanik1996recent}. \ansnew{We verified this problem by comparing the data to ice-coverage maps published by the Arctic and Antarctic Research Institute (AARI), who compiles not only satellite information but also those from coastal stations and ship reports \cite{AARIMAPreference}. Examples of such artefacts are given in Fig.~\ref{Fig_anomalies_and_resto}(a).}

To improve the quality of the OSI SAF data upon assimilation we compute a `reliability coefficient' $w$ for each grid point  based on the distance to the nearest point on land.
Fig.~\ref{Fig_anomalies_and_resto}(b) shows the values we obtained for $w$; these are stored as a complement to the model grid data.

\ansnew{As for ice thickness data, we used the CS2SMOS dataset \cite{ricker2017weekly}. 
It combines data from the SMOS and CryoSat-2 ice-thickness datasets, each one represents
thin ($<$ 1 m) and thick ice with different errors. As pointed out in \cite{mu2018improving}, joint assimilation of these datasets increases ice simulation quality. The CS2SMOS dataset is provided with weekly time resolution; therefore, values were linearly interpolated in time for usage in our simulations. This dataset also contains the uncertainty associated 
to each thickness value, which enables us to apply the restoring scheme only when model and satellite ice-thickness data disagree significantly --- i.e., when model results are outside the confidence interval of observations.}

Once satellite data are made available on the same grid as the ice-model state variables, a simple non-statistical state-restoring scheme can be used to force the model state to conform to observations \cite{griffies2009coordinated}.
    
\ansnew{While direct methods to modify ice-state parameters (e.g. \cite{mu2018improving}) can be sufficient in post-processing pipelines, indirect methods such as the adjoint-based method described in \cite{koldunov_adjoint} better maintain data self-consistency during
simulations. We therefore adopt an indirect, flux-based restoring method.} 

We compare ice concentration and thickness fields against the corresponding satellite observations, and the ice state is controlled using an additional feedback term in the heat flux equations. The melting of excess ice is forced by increasing the atmosphere-to-ice heat flux, and the formation of missing ice is forced by increasing the ice-to-ocean non-solar heat flux as follows:

\ansnew{
\begin{equation}
\label{eq_damp}
\begin{array}{rcll}
Q_{sr}^*&=&Q_{sr} + w\,\widetilde{Q}_{sr}, & \text{if there is excess ice, and} \\
Q_{ns}^*&=&Q_{ns} + w\,\widetilde{Q}_{ns}, & \text{if there is missing ice.} 
\end{array}
\end{equation}
}

\begin{figure} [ht!]
  \center
  \includegraphics [scale=0.9] {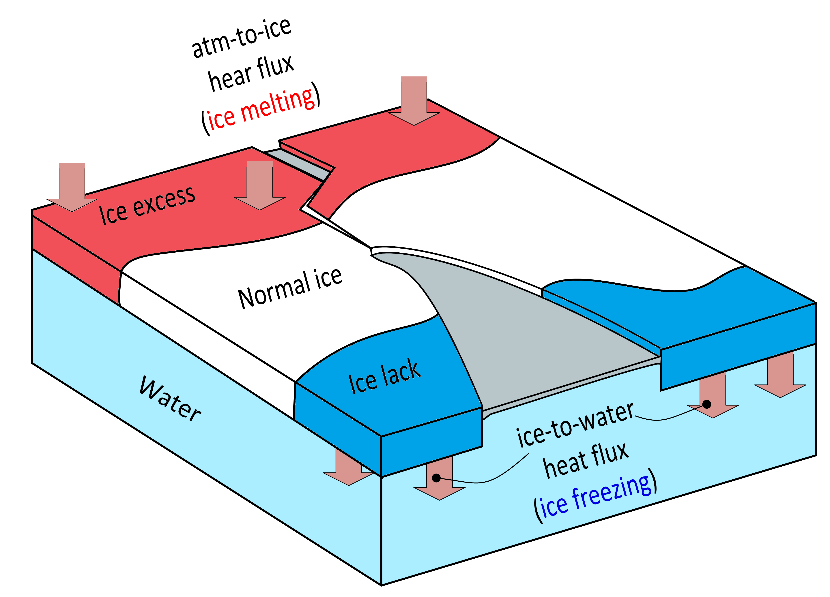}
  \caption{Using observational data to drive the damping of 
  the sea-ice state through modified incoming and outgoing heat fluxes.}
  \label{fig_to_be_removed}
\end{figure}

In Eq.\ref{eq_damp}, $Q_{sr}$ and $Q_{ns}$ are the heat fluxes before this forcing is applied, the subscripts $sr$ and $ns$ refer to solar (atmosphere-to-ice) and non-solar (ice-to-ocean), respectively (see Fig.~\ref{fig_to_be_removed}). $\widetilde{Q}$ is the damping term, and $w$ is the reliability coefficient. The observation-driven computation of $\widetilde{Q}_{sr}$ and $\widetilde{Q}_{ns}$ allows real-time damping of the model state.

Ice observations reflect an averaged value over an area. For each grid cell, we interpolate the satellite data to obtain the observed concentration $C_{obs}$ and thickness $H_{obs}$, with associated $\pm\sigma_C$ and $\pm\sigma_H$ confidence intervals derived from satellite total error. NEMO-LIM3, however, stores ice concentration and thickness in the form of several values per cell, each value associated with one of $L$ ice-thickness categories \cite{lim36refguide}. A single value representative of all categories must be computed for comparison with satellite data. For each cell, the model values for total ice concentration and thickness are given by 

\ansnew{
\begin{equation}
\label{ice_conc_thick_mean}
  \overline C = \sum_L C_\ell, \quad\quad\text{and} \quad\quad
  \overline H = \sum_L C_\ell H_\ell\quad\text{.}
\end{equation}
For each grid cell, we denote the difference between model state
and observed values as

\begin{equation}
\label{ice_conc_thick_diff}
\Delta C =\overline C - C_{obs} \quad\text{and}\quad 
\Delta H =\overline H - H_{obs}.
\end{equation}
\smallskip

In the event that either $C_{obs}$ or $H_{obs}$ is not available,
the corresponding $\Delta C$ or $\Delta H$ is set to zero.

The four $\alpha$ values ($\alpha_{m}^C$, $\alpha_{m}^C$, $\alpha_{m}^H$, $\alpha_{f}^H$) are positive coefficients for ice concentration- and thickness restoring: the subscripts $m$ and $f$ correspond to `melting' and `formation', and the superscripts $C$ and $H$ indicate whether they are associated to $\Delta C$ or $\Delta H$, i.e., the values are different. Given that $\Delta C$ is a fraction and $\Delta H$ is in meters, the coefficients $\alpha_{m,f}^C$ are given in $\lbrack W/m^2\rbrack$ and $\alpha_{m,f}^H$ are given in $\lbrack W/m^3\rbrack$ (considering $\widetilde{Q}$ is given in $\lbrack W/m^2\rbrack$)

Using these quantities, we define the damping terms of Eq.\ref{eq_damp} as

\begin{equation}
\label{eq_damp_term_sr}
\widetilde{Q}_{sr} =\left\lbrace
\begin{array}{ll}
  \alpha_m^C\,\Delta C, & \text{for}\quad\Delta C > \sigma_C, \\
  \phantom{.}\\
  \alpha_m^H\,\Delta H, & \text{for}\quad|\Delta C| \le \sigma_C \quad\text{and}\quad \Delta H > \sigma_H, \\
  \phantom{.}\\
  0, & \text{otherwise, and}
\end{array}
\right.\\
\end{equation}

\begin{equation}
\label{eq_damp_term_ns}
\widetilde{Q}_{ns} =\left\lbrace
\begin{array}{ll}
  -\alpha_f^C\,\Delta C, & \text{for}\quad\Delta C < -\sigma_C, \\
  \phantom{.}\\
  -\alpha_f^H\,\Delta H, & \text{for}\quad|\Delta C| \le \sigma_C \quad\text{and}\quad \Delta H < -\sigma_H, \\
  \phantom{.}\\
  0, & \text{otherwise.}
\end{array}
\right.
\end{equation}

Substituting Eqs.\ref{eq_damp_term_sr}-\ref{eq_damp_term_ns} 
in Eq.\ref{eq_damp}, 
the corrected heat flux equations for ice melting and formation become

\begin{equation}
\label{eq_damp2_sr}
Q_{sr}^*=\left\lbrace
\begin{array}{ll}
  Q_{sr} + w\left(\alpha_m^C\,\Delta C\right), & \text{for}\quad\Delta C > \sigma_C, \\
  \phantom{.}\\
  Q_{sr} + w\left(\alpha_m^H\,\Delta H\right), & \text{for}\quad|\Delta C| \le \sigma_C \quad\text{and}\quad \Delta H > \sigma_H, \\
  \phantom{.}\\
  Q_{sr}, & \text{otherwise, and}
\end{array}
\right.
\end{equation}

\begin{equation}
\label{eq_damp2_ns}
Q_{ns}^* =\left\lbrace
\begin{array}{ll}
  Q_{ns} - w\left(\alpha_f^C\,\Delta C\right), & \text{for}\quad\Delta C < -\sigma_C, \\
  \phantom{.}\\
  Q_{ns} - w\left(\alpha_f^H\,\Delta H\right), & \text{for}\quad|\Delta C| \le \sigma_C \quad\text{and}\quad \Delta H < -\sigma_H, \\
  \phantom{.}\\
  Q_{ns}, & \text{otherwise.}
\end{array}
\right.
\end{equation}
\smallskip

}

\begin{figure} [ht!]
\center
\includegraphics [scale=0.4] {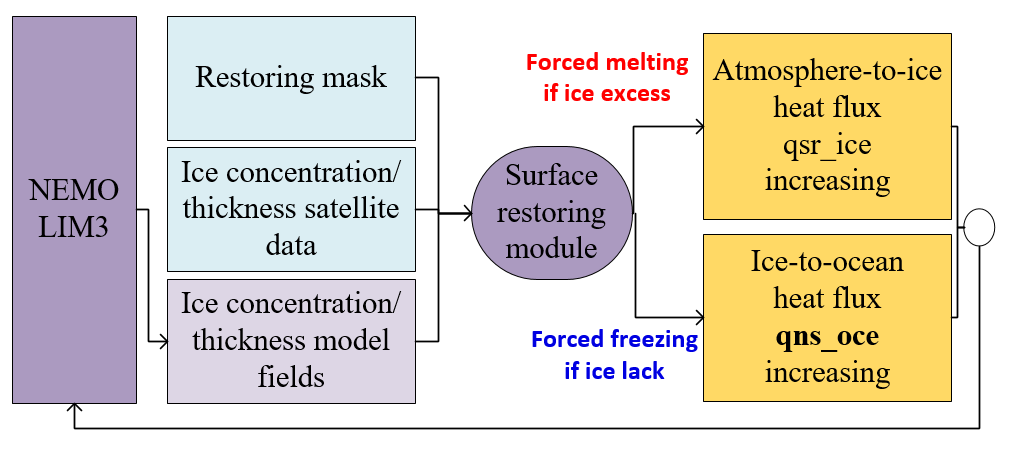}
\caption{Flowchart for ice-state restoring, as
implemented in the \textit{sbc\_ssr\_ice} module.}
\label{fig7}
\end{figure}

NEMO already implements a method for restoring sea surface temperature (SST) and salinity (SSS), based on correction of the heat and fresh-water fluxes, \ansnew{which similarly adds damping terms to the corresponding equations. We extended the corresponding routines to incorporate the formulation outlined above. A diagram of the custom ice restoring scheme is presented in Fig.~\ref{fig7}.}

\subsection{Estuary temperature stabilisation}
\label{sec53}

\begin{figure} [ht!]
\center
\includegraphics [scale=0.4] {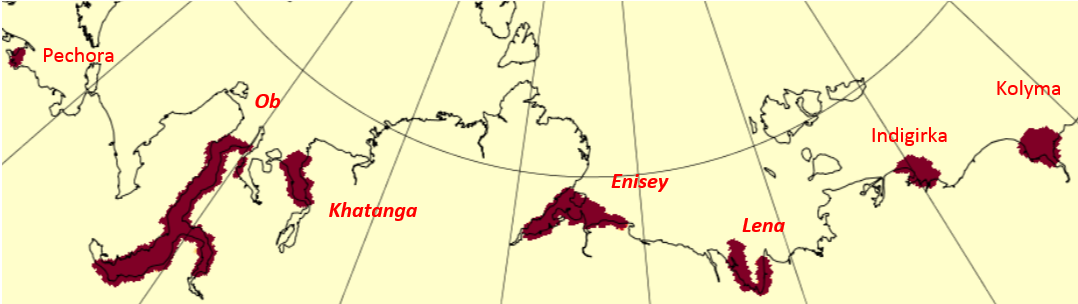}
\caption{The estuary regions for main Russian Arctic rivers.}
\label{fig_river_mask}
\end{figure}

The NEMO model allows setting the river runoff parameters as part of the 
surface boundary conditions. Temperature, salinity and runoff are specified for grid cells corresponding to estuaries as time-dependent variables. These areas of the domain are depicted in Fig.~\ref{fig_river_mask}: \ansnew{this values was obtained from the global NEMO configuration ORCA025\cite{bourdalle2006climatology} input dataset, as well as the corresponding runoff values \cite{molines2012definition}. Temperature values were taken from multi-year dataset for the Russian Arctic rivers \cite{WaterCadastre}.}

Initial temperature and salinity conditions for coastal Arctic regions obtained from ocean reanalyses are not precise enough due to the relatively low resolution. The default implementation in NEMO relies on the quality of the initial data: it includes damping terms in the temperature and salinity flux equations to simulate estuary flow. As a consequence of the poor coastal data quality, it requires a long spin-up time to stabilise the coastal thermohaline system, i.e. to balance the temperature and salinity in the river estuary and nearby ocean area. \ansnew{To circumvent this issue and accelerate the convergence of the thermohaline spin-up, we replace the damping scheme with direct overriding of model state with temperature and salinity values from the data described above.}

\begin{figure} [ht!]
\center
\includegraphics [scale=0.45] {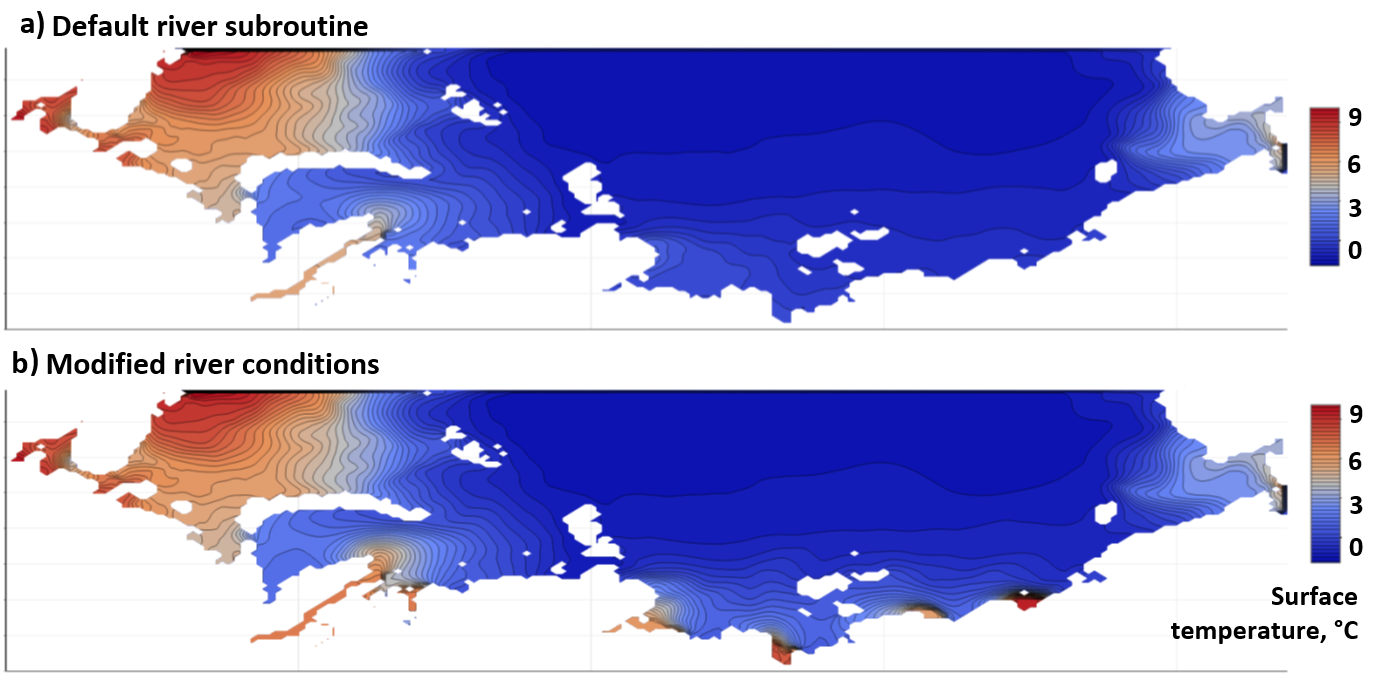}
\caption{Comparison of summer-averaged surface temperature fields obtained 
after a 6-month spin-up with a) the modified estuary
scheme b) the default scheme}
\label{river-temp-comparison}
\end{figure}

A set of one-year experiments was conducted to validate the effects of the modifications applied to the estuary scheme. Fig.~\ref{river-temp-comparison}(b) shows that the modified scheme is successful in representing the influence of rivers on the sea surface temperature inside and near their estuaries, even with such a short spin-up time. The same is not observed in Fig.~\ref{river-temp-comparison}(a): the default scheme would require a much longer spin-up to reach a similar state.

\subsection{Validation of the implemented ice restoring}
\label{sec54}

Numerical experiments were conducted to verify the proposed ice-restoring
approach, confirming its reduced spin-up time
and increased simulation-result quality.

Each experiment followed the procedure described in Sec.\ref{sec51}.
In the coarse-grid spin-up runs,
ice state was initialised using the default LIM3 configuration,
which calculates initial ice concentrations from sea-water temperatures 
obtained from the GLORYS2V4 reanalysis, as well as 
initial temperatures and salinity state for the ocean.
      
\begin{figure} [ht!]
\center
\includegraphics [scale=0.5] {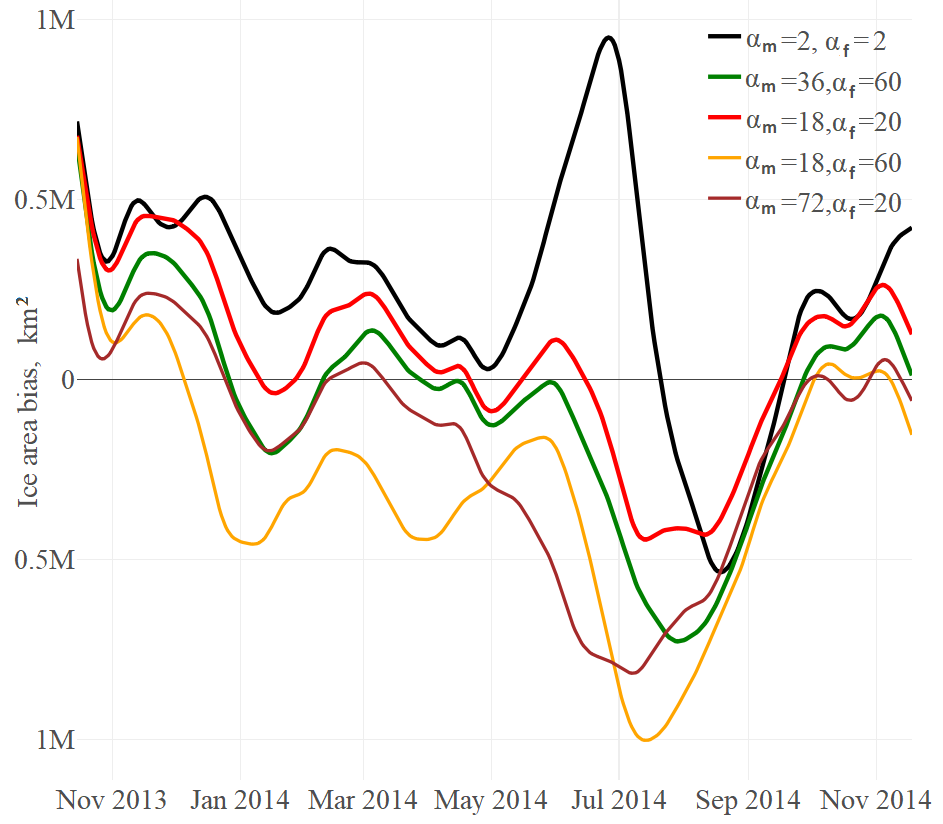}
\caption{Ice-coverage bias for different values of restoration coefficients $\alpha_m^C$ and
$\alpha_f^ C$ on the coarse grid.}
\label{fig71}
\end{figure}

The $\alpha$ coefficients from Eqs.\ref{eq_damp2_sr}-\ref{eq_damp2_ns}
were determined in a series of one-year experiments.
The total ice cover bias against satellite observations for 
different melting 
and freezing ice-coverage coefficients is presented in Fig.~\ref{fig71}, 
and the 
time-averaged values of corresponding error metrics are presented in 
Table \ref{Tab_weights_results}. 
    
\begin{table}[h!]
\centering
\caption{Total ice coverage error metrics for model runs with different values of restoring coefficients $\alpha_m^C$ and $\alpha_f^ C$ on the coarse grid. The boldface numbers indicate the best coefficients set found.}
\label{Tab_weights_results}
\begin{tabular}{|c|c|c|c|c|}
\hline
$\vphantom{\text{\Large O}}\alpha_f^C$ & $\alpha_m^C$ & Bias & RMSE & MAE \\ 
$\lbrack W/\text{m}^2\rbrack$ & 
$\lbrack W/\text{m}^2\rbrack$ & 
$\lbrack10^3 \; \text{km}^2\rbrack$ & 
$\lbrack10^3 \; \text{km}^2\rbrack$ & 
$\lbrack10^3 \; \text{km}^2\rbrack$ \\ \hline
2 & 2 & 239          & 400  & 338 \\ \hline
36 & 60 & -73               & 306  & 230 \\ \hline
\textbf{18} & \textbf{20} & \textbf{58}    & \textbf{265}  & \textbf{213} \\ \hline
18 & 60 & -290               & 438  & 348 \\ \hline
72 & 20 & --195               & 368  & 267 \\ \hline

\end{tabular}
\end{table}

From these results, we chose the coefficient pair
$\alpha_f^ C = 18$ and $\alpha_m^C=20$; then $\alpha_{f,m}^H$ were
obtained in a similar procedure (and set to 40 and 56 respectively).

\begin{figure} [ht!]
\center
\includegraphics [scale=0.60] {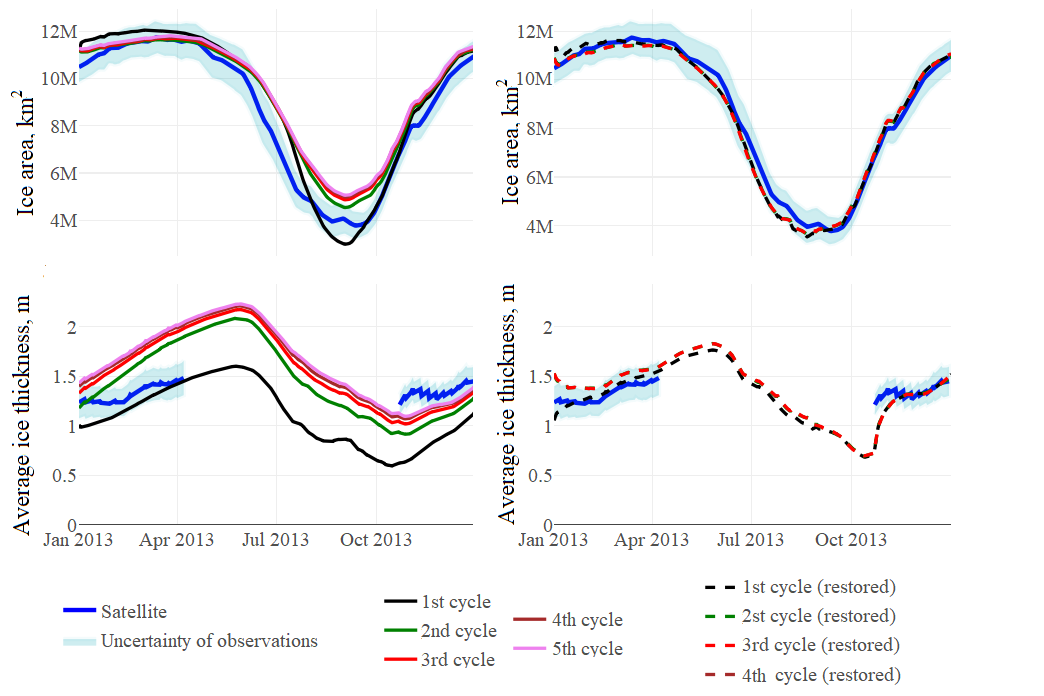}
\caption{Total ice coverage (top) and \ansnew{average} ice thickness (bottom)
on the coarse grid (full Arctic Ocean) for different runs,
without ice-restoring (left) and with it (right).}
\label{fig_coarse_res}
\end{figure}

We then performed comparative experiments both with and without our
ice-restoring scheme over the date range
from 01.01.2013 to 31.12.2013.
Two-stage spin-up was performed as described in Sec.\ref{sec51}.
The ice area and \ansnew{average thickness}
metrics for the coarse-grid configuration 
are presented in Fig.~\ref{fig_coarse_res}. 

\begin{figure} [ht!]
  \center
  \includegraphics [scale=0.65] {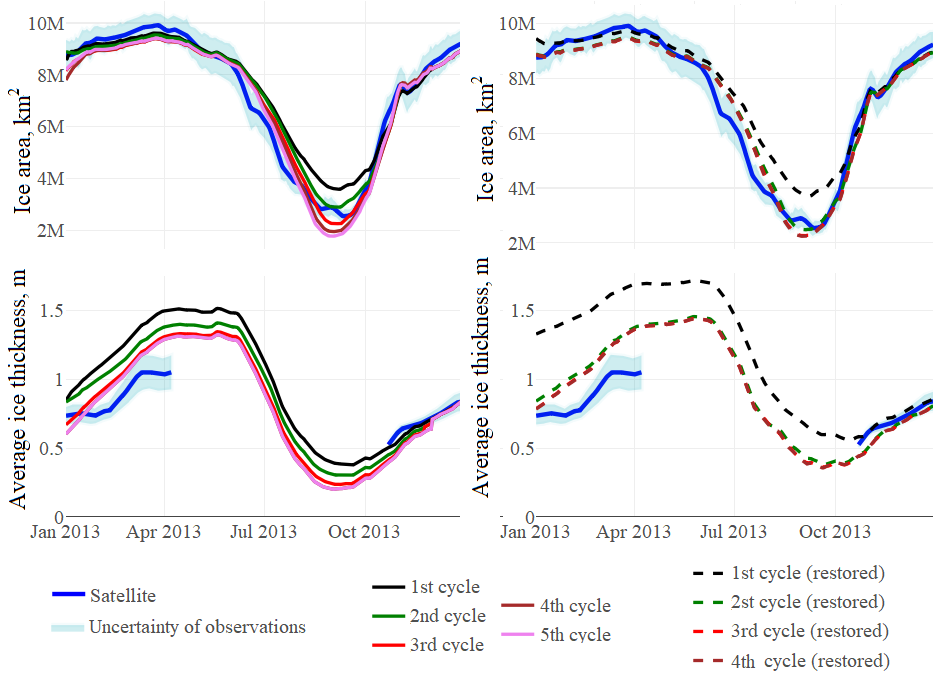}
        \caption{Total ice coverage (top) and \ansnew{average} ice thickness (bottom)
on the fine grid (full Arctic Ocean) for different runs,
without ice-restoring (left) and with it (right).}
  \label{fig_nested_res}
\end{figure}

It can be seen that the restoring-based model provides results 
closer to satellite observations than the non-restored model
in the ice melting period (May-September).
Corresponding plots for the fine-grid spin-up are presented
in Fig.~\ref{fig_nested_res}

\begin{figure} [ht!]
\center
\includegraphics [scale=0.53] {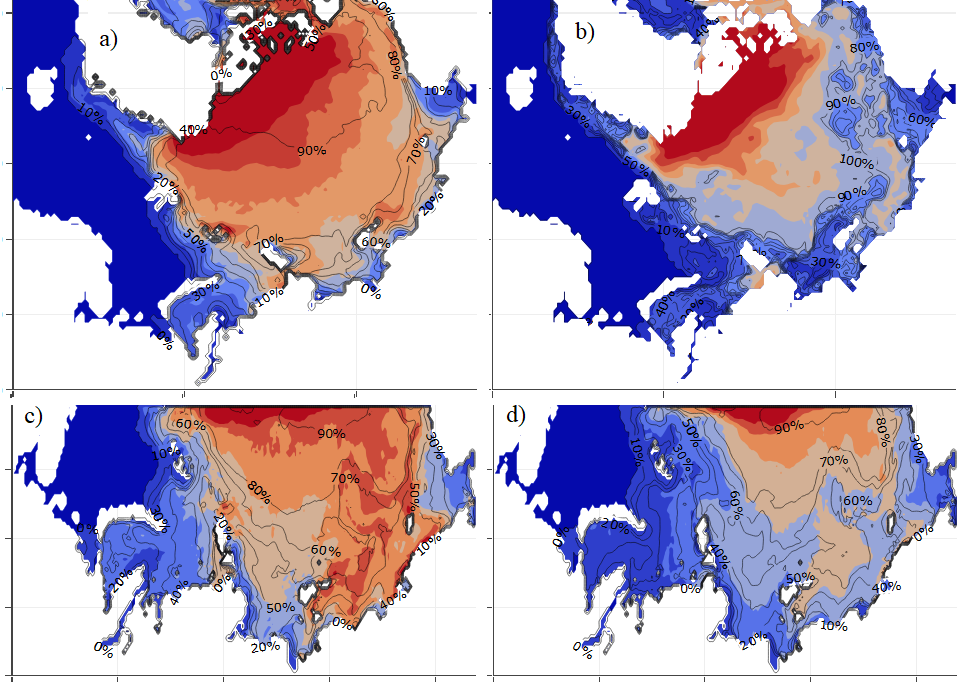}
\caption{Visualisation of summer-averaged ice fields after
5-year spin-up without ice-restoring (left) and 
3-year spin-up with ice-restoring (right).
The top figures show coarse-grid results (full Arctic Ocean)
of the primary spin-up,
and the bottom figures show fine-grid results (Russian Arctic seas)
of the secondary spin-up.
The colour scale corresponds to ice thickness, and 
contour lines correspond to ice concentration (0-100\%).}
\label{fig_fields_viz}
\end{figure}

Both coarse- and fine-grid spin-up runs stabilise after one or two
cycles when ice-restoring is applied, otherwise the same procedure takes
at least four cycles. Table \ref{Tab_Summary} presents error metrics
for both resolutions, with and without our ice-restoring scheme:
root mean-square error and mean absolute error were computed 
for each cell over the period 01.06.2013-31.09.2013 and 
summed over the whole domain. The corresponding time-averaged concentration
and thickness are shown in Fig.~\ref{fig_fields_viz}. 

Table \ref{Tab_Summary} shows that errors for the coarse-grid without the ice-restoring scheme actually increase as the solution stabilises. In addition to reaching lower values for the error metrics, the results of runs with ice-restoring are qualitatively different, as shown in Fig.~\ref{fig_fields_viz}. Taken together, these results confirm that the proposed scheme reaches a better solution at a lower cost.

\begin{table}[ht!]
\centering
\caption{Error metrics for total ice coverage and \ansnew{average} ice thickness after each cycle of the spin-up experiments. The boldface numbers indicate the converged solution for the restored runs.}
\label{Tab_Summary}
\begin{tabular}{|c|c|c|c|c|c|c|c|c|}
\hline
 & \multicolumn{4}{c|}{Non-restored} & \multicolumn{4}{c|}{Restored} \\ \cline{2-9} 
\multirow{-2}{*}{Cycle} & \begin{tabular}[c]{@{}c@{}}Area\\   RMSE\\   ($10^6 \; \text{km}^2$)\end{tabular} & \begin{tabular}[c]{@{}c@{}}Thick. \\ RMSE\\   m\end{tabular} & \begin{tabular}[c]{@{}c@{}}Area\\   MAE\\    ($10^6 \; \text{km}^2$) \end{tabular} & \begin{tabular}[c]{@{}c@{}}Thick. \\ MAE\\   m\end{tabular} & \begin{tabular}[c]{@{}c@{}}Area\\   RMSE\\   ($10^6 \; \text{km}^2$)\end{tabular} & \begin{tabular}[c]{@{}c@{}}Thick. \\ RMSE\\   (m)\end{tabular} & \begin{tabular}[c]{@{}c@{}}Area\\   MAE\\   ($10^6 \; \text{km}^2$)\end{tabular} & \begin{tabular}[c]{@{}c@{}}Thick. \\ MAE,\\   m\end{tabular} \\ \hline
\multicolumn{9}{|c|}{Coarse-grid (14km)} \\ \hline
1 & 0.69 & 0.34 & 0.57 & 0.28 & 0.34 & 0.09 & 0.27 & 0.06 \\ \hline
2 & 0.76 & 0.28 & 0.59 & 0.26 & \textbf{0.33} & \textbf{0.13} & \textbf{0.27} & \textbf{0.11} \\ \hline
3 & 0.89 & 0.31 & 0.70 & 0.29 & 0.33 & 0.13 & 0.28 & 0.11 \\ \hline
4 & 0.94 & 0.33 & 0.76 & 0.30 & 0.33 & 0.13 & 0.28 & 0.11 \\ \hline
\multicolumn{9}{|c|}{Fine-grid (5km)} \\ \hline
1 & 0.38 & 0.26 & 0.28 & 0.20 & 0.43 & 0.48 & 0.30 & 0.37 \\ \hline
2 & 0.29 & 0.19 & 0.23 & 0.16 & 0.26 & 0.19 & 0.20 & 0.16 \\ \hline
3 & 0.27 & 0.13 & 0.23 & 0.11 & \textbf{0.25} & \textbf{0.17} & \textbf{0.20} & \textbf{0.14} \\ \hline
4 & 0.28 & 0.12 & 0.24 & 0.10 & 0.25 & 0.17 & 0.20 & 0.14 \\ \hline
\end{tabular}
\end{table}

\section{Conclusion}
\label{sec6}

\ansnew{We have implemented a two-scale NEMO regional configuration 
for the Russian Arctic seas. The primary goal with this configuration is to run long-term hindcasts at high spatio-temporal resolution, as required to obtain dataset for environmental risk assessment.} The main advantages of the innovations presented in this article are:

\begin{itemize}
\item obtaining high-resolution metocean fields for subdomains of
a coarse-grid domain at a lower cost compared to a high-resolution 
simulation of the whole domain;
\item no restrictions are imposed on the coarse-grid for the full domain
by any choice of fine-grid and sub-region;
\item the ability to run the model on the coarse and fine grids separately,
allowing for better usage of computational resources;
\item transparency at the coupling stage, allowing intermediate data
to be examined and modified;
\item increased quality of regional ice modelling due to the 
implementation of the ice drift boundary condition; and
\item reduced spin-up time and increased quality of the both coarse- and fine-grid configurations made possible by the ice-restoring scheme. 
\end{itemize}

The method and implementation developed in this work can also be used in other contexts. For example, the OBC procedures for tracers and ice drift are suitable for various regional configurations with long open boundaries, and ice restoring is applicable for ice-covered domains in cases where observational data are available. Some parts of the developed configuration can also be used in an `online' coupled version, either to simplify the computational pipeline or improve result quality.

The modified NEMO source code described in this article
is publicly available \cite{nemo-multigrid}.

\section{Acknowledgements}
\label{sec7}

This research is financially supported by The Russian Scientific Foundation.

The research is carried out using the equipment of the shared research facilities of HPC computing resources at Lomonosov Moscow State University.

The CryoSat-2/SMOS data from 2013.01.01 to 2014.12.31 are provided by the \\
http://www.meereisportal.de (grant: REKLIM-2013-04). 

We are further indebted to Gustavo Hime, who helped us bring the quality of 
this manuscript to publication standards.

\bibliographystyle{elsarticle-num}
\bibliography{nemo-article-black}

\appendix

\section{Grid distortion computation} \label{App_grids}
\setcounter{table}{0}
A common approach to the orthogonal curvilinear grid generation is to 
exploit properties of conformal map projections 
\cite{murray1996explicit,bentsen1999coordinate}. 
We denote a mapping $F_p:(x, y)\mapsto(\phi, \lambda)$ from 
Cartesian to geographical coordinates associated 
with a projection $p$ as:

\begin{equation}
F_p(x, y) = R_p(P_p(x, y))\text{,}
\end{equation}
where $P_p:(x, y)\mapsto(\phi_p, \lambda_p)$ is the 
inverse projection transformation (i.e. plane-to-sphere) for the case 
of a spherical Earth approximation, and 
$R_p:(\phi_p, \lambda_p)\mapsto(\phi, \lambda)$ denotes a sequence of 
rotations of the latitude-longitude grid that shifts the reference point of 
the projection to the central point of the computational domain. 
In the case of a conformal projection, 
the scale factors $e_1(\phi, \lambda)$ and $e_2(\phi, \lambda)$ 
required by NEMO \cite{NemoGuide} are equal to each other and can 
be calculated by using a corresponding analytical expression for scale 
distortions 
$h_p:(\phi_p, \lambda_p)\mapsto\mathbb{R}$ as shown in \cite{snyder1987map}:

\begin{equation}
e_1(\phi, \lambda) = e_2 (\phi, \lambda) = \dfrac{s}{h_p[R_p^{-1}(\phi, \lambda)]}\text{,}
\end{equation}
where $s$ is the Cartesian grid spacing.

We tested the Stereographic, Mercator and Lambert Conformal Conic 
projections to identify which one introduced the least distortion 
for our particular Cartesian grid.
Each projection is parameterised by a vector $\boldsymbol{\phi}_{ts}$ of 
latitudes of true scale (the length of the vector depends on the projection).
For a given projection $p$, mean distortion $d_p$ for a 
particular Cartesian grid can be written as

\begin{equation}
\label{mean_dist}
d_p(\boldsymbol{\phi}_{ts}) = \dfrac{s}{N\times M} \sum_i^N \sum_j^M{\left|\dfrac{1}{h_{p(\boldsymbol{\phi}_{ts})}\left[P_{p(\boldsymbol{\phi}_{ts})}(x_{ij}, y_{ij})\right]} - 1\right|}\text{,}
\end{equation}
where $N$ and $M$ are the grid size. For each projection, 
we minimise the target function Eq.\ref{mean_dist} 
to find values of $\boldsymbol{\phi}_{ts}$ that 
produce the minimal mean linear distortion in the grid points of a 
particular Cartesian grid. 
The relative distortion values for each projection are given in 
Table \ref{TabGridDistortions}.

\begin{table}[h!]
\centering
\caption{Comparison of Arctic regional grid distortions for 
different projections}
\label{TabGridDistortions}
\begin{tabularx}{\textwidth}{||l||Y|Y||Y|Y||}
\hline
\multicolumn{1}{||c||}{\multirow{2}{*}{Projection}} & \multicolumn{2}{c||}{Distortion in all grid points} & \multicolumn{2}{c||}{Distortion in sea points} \\ \cline{2-5} 
\multicolumn{1}{||c||}{} & Mean & Max & Mean & Max \\ \hline
Stereographic & 2.0\% & 3.7\% & 1.9\% & 3.7\% \\ \hline
Mercator & 3.2\% & 4.3\% & 2.7\% & 4.3\% \\ \hline
Lambert Conformal Conic & 3.1\% & 5.7\% & 2.5\% & 5.7\% \\ \hline
 \end{tabularx}
\end{table}

\section{Vector notation} \label{App_vectors}
\setcounter{figure}{0}

\begin{figure} [ht!]
  \center
  \includegraphics [scale=0.4] {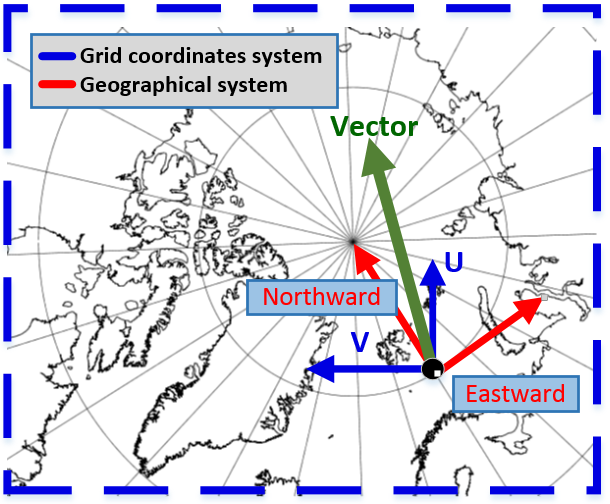}
  \caption{A visualisation of the difference between coordinate systems for 
  vector variables.}
  \label{fig_vect_notation_diff}
\end{figure}

Fig.~\ref{fig_vect_notation_diff} illustrates the notion of
geographical and local coordinate systems.

Means to rotate the U and V components of input 
vector fields from the geographical system to a local coordinate system
were 
implemented as the \textit{rot\_rep} subroutine in the 
\textit{geo2ocean.F90} source file. 
It also contains the \textit{rot\_rep\_point} 
subroutine, which processes just one grid point.
Further modifications were introduced to \textit{bdytides.F90},
which contain code related to tidal currents, and to
\textit{bdydta.F90}, 
which contain code related to barotropic and baroclinic currents:
these modifications implement the pre-processing of open boundary conditions,
rotating vectors using the routines described above.

Output vector variables 
--- e.g. ocean currents and ice drift --- 
are computed in the local grid system, and must be transformed back into
the geographical system before any further post-processing.
Modifications were introduced to \textit{diawri.F90},
where calls to \textit{rot\_rep} performs the required transformation.
These source code changes allow all input and output datasets to be 
pre-/post-processed in geographical notation.

\section{Tracer and ice drift open boundary conditions} 
\label{App_tracer_boundary}

Ocean boundary data can be divided into inflow and outflow 
zones \cite{OBCdiscussion,OBCreference}, i.e. zones with positive and 
negative phase speed. 
For any tracer or a flux component $\varphi$, phase speed is usually defined as

\begin{equation}
\label{phase_vel}
c_\varphi=
\frac{\sfrac{\partial \phi}{\partial t}}{\sfrac{\partial\phi}{\partial x}}.
\end{equation}

However, there are different numerical schemes for computing 
phase speed.  For consistency, we adopt the one used in
NEMO:

\begin{equation}
\label{phase_num}
   c_\varphi=\frac{\varphi_{B+1}^b-\varphi_{B+1}^a}{\varphi^a_{B+1}-\varphi^a_{B+2}}
\end{equation}

The upper index in Eq.\ref{phase_num} is the time grid index, 
i.e. $\varphi^a$ is the variable ``after" the integration time and 
$\varphi^b$ is the variable ``before" the integration time. 
The lower index is the boundary position index, 
i.e. $\varphi_{B+1}$ is the variable at the ``boundary plus one" cell.

The widely used tracer OBC schemes implemented in the NEMO model 
are FRS (flow relaxation scheme) and Orlanski conditions. 
FRS \cite{OBCreference} has a strong nudging effect, 
which decreases with the increasing distance from the boundary. 
The standard FRS scheme is written as:

\begin{equation}
\label{frs_scheme}
   \bar{\varphi}= \alpha  \varphi_e + (1-\alpha) \varphi_i
\end{equation}

Eq.\ref{frs_scheme} uses notation from \cite{FRSreference}:
$\varphi_e$ is the external data value, $\varphi_i$ is the internal model 
value before the OBC updating procedure, 
$\bar{\varphi}$ is the model value after the OBC updating procedure. 
The relaxation parameter $\alpha$ varies in the range from 0 to 1 in the 
boundary outer normal direction. In NEMO, 
$\alpha=1-\tanh{\frac{d}{2}}$, { where} $d$ is the distance from the open 
boundary. 
It should be noted that the FRS scheme could also be used for other model 
components at the boundary (e.g., ice drift).
We used the FRS scheme described above in the scope of this work.

In NEMO, the Orlanski radiation scheme is available as OBC for tracers. 
It has the following implicit form:

\begin{equation}
\label{orlanski_implicit}
   \frac{\partial \varphi}{\partial t}+c_\varphi \frac{\partial \varphi}{\partial n}=0
\end{equation}
where phase velocity $c_\varphi$ is given in Eq.\ref{phase_vel}.

Applying Eq. \ref{orlanski_implicit} to Eq. \ref{phase_num} gives
the OBC updating procedure

\begin{equation}
   \varphi_{B}^a=\varphi_{B}^b-\bar{c}_\varphi (\varphi_{B+1}^a-\varphi_{B}^a).
\end{equation}

In the literature \cite{OBCreference}, the full tracer 
velocity is defined as the sum of phase velocity $c_T$
upstream velocity $u$ 
(the ocean velocity in the direction, normal to the boundary).
Accordingly, the advection scheme for tracers is given by

\begin{equation}
\label{adv_explicit}
   \frac{\partial T}{\partial t}+(c_T+u)\frac{\partial T}{\partial n}=0,
\end{equation}

which can be discretised using finite differences as:

\begin{equation}
\label{adv_implicit}
   T_{B}^a=T_{B}^b-(\bar{c}_T+u_B^b) (T_{B+1}^a-T_{B}^a)
\end{equation}

The scheme in Eq.\ref{adv_implicit} is the 
advection scheme (ADV) in this article.

\section{Summary of changes in the NEMO source code} \label{App_code_modifications}

\setcounter{table}{0}

The list of most significant modifications that were made to the
NEMO ocean model 3.6 (build 7873) source code is presented in 
Tab.\ref{tab_sources}.

\begin{table}[ht!]
\caption{Location of changes in the source code of NEMO 3.6 build 7873}
\label{tab_sources}
\begin{tabular}{|l|l|l|}
\hline
Section & Changes & File \\ \hline
  Section 3      &  Input-output vector variable notation modified   &    geo2ocean.F90  \\ 
          &     to eastward-northward      &   bdydta.F90  \\ 
           &       &     bdytides.F90  \\ 
           &      &    diawri.F90  \\ \hline
    Section 4.1    &    ADV(+FRS) schemes   &   bdylib.F90   \\\hline
        Section 4.1    &    ADV(+FRS) subroutine calls    &   bdytra.F90   \\\hline
        Section 4.1    &    ADV(+FRS) choice output in ocean.output    &   bdyini.F90   \\\hline
        Section 4.2    &    Drift FRS choice output in ocean.output    &   bdyini.F90   \\
                        &    Drift boundary data switch for ice drift    &     \\\hline
        Section 4.2    &    Ice drift FRS scheme implementation    &   bdyice\_lim.F90   \\\hline
        Section 4.2    &    Ice drift boundary conditions switch  &   bdy\_oce.F90   \\\hline
        Section 5.2 & Ice concentration and thickness \eng{restoring} implementation & sbcssr.F90 \\\hline
        Section 5.3    &    River mounth temperature stabilisation implementation    &   sbcssr.F90   \\\hline
        &    &   bdytra.F90   \\\hline
\end{tabular}
\end{table}

The modified NEMO source code can be obtained from the public repository 
\cite{nemo-multigrid}.

\end{document}